\documentclass[5p,times,twocolumn]{elsarticle}
\usepackage{lineno,hyperref}
\usepackage{subcaption}
\usepackage{tikz}
\usepackage{pgfplotstable}
\usepackage{pgfplots}
\usepackage{multirow}
\usepackage{float}
\usepackage{textcomp}
\usepackage{amsmath,amsfonts,amssymb,bbm}	
\usepackage{multicol}
\usepackage{siunitx}
\usepackage{booktabs}
\usepackage{soul}

\usepackage[section]{placeins}

\modulolinenumbers[5]

\biboptions{authoryear}

\usepackage{multirow} 
\usepackage{booktabs} 
\usepackage{graphicx} 
\usepackage{placeins} 
\usepackage{hyperref}
\usepackage{rotating} 

\usepackage{blindtext}

\begin{document}

\begin{frontmatter}

\title{Sinogram upsampling using Primal-Dual UNet for undersampled CT and radial MRI reconstruction}

\author[1,2,4]{Philipp Ernst\corref{equalcontribution}}
\author[1,2,3,4,5]{Soumick Chatterjee\corref{equalcontribution}}
\author[4,6]{Georg Rose}
\author[3,4,7,8]{Oliver~Speck}
\author[1,2,8]{Andreas N{\"u}rnberger}

\cortext[equalcontribution]{P. Ernst and S. Chatterjee have Equal Contribution}

\address[1]{Faculty of Computer Science, Otto von Guericke University Magdeburg, Germany}
\address[2]{Data and Knowledge Engineering Group, Otto von Guericke University Magdeburg, Germany}
\address[3]{Biomedical Magnetic Resonance, Otto von Guericke University Magdeburg, Germany}
\address[4]{Research Campus STIMULATE, Otto von Guericke University Magdeburg, Germany}
\address[5]{Genomics Research Centre, Human Technopole, Milan, Italy}
\address[6]{Institute for Medical Engineering, Otto von Guericke University Magdeburg, Germany}
\address[7]{German Centre for Neurodegenerative Disease, Magdeburg, Germany}
\address[8]{Centre for Behavioural Brain Sciences, Magdeburg, Germany}

\begin{abstract}
Computed tomography (CT) and magnetic resonance imaging (MRI) are two widely used clinical imaging modalities for non-invasive diagnosis. However, both of these modalities come with certain problems. CT uses harmful ionising radiation, and MRI suffers from slow acquisition speed. Both problems can be tackled by undersampling, such as sparse sampling. However, such undersampled data leads to lower resolution and introduces artefacts. Several techniques, including deep learning based methods, have been proposed to reconstruct such data. However, the undersampled reconstruction problem for these two modalities was always considered as two different problems and tackled separately by different research works. This paper proposes a unified solution for both sparse CT and undersampled radial MRI reconstruction, achieved by applying Fourier transform-based pre-processing on the radial MRI and then finally reconstructing both modalities using sinogram upsampling combined with filtered back-projection. The Primal-Dual network is a deep learning based method for reconstructing sparsely-sampled CT data. This paper introduces Primal-Dual UNet, which improves the Primal-Dual network in terms of accuracy and reconstruction speed. The proposed method resulted in an average SSIM of 0.932\textpm0.021 while performing sparse CT reconstruction for fan-beam geometry with a sparsity level of 16, achieving a statistically significant improvement over the previous model, which resulted in 0.919\textpm0.016. Furthermore, the proposed model resulted in 0.903\textpm0.019 and 0.957\textpm0.023 average SSIM while reconstructing undersampled brain and abdominal MRI data with an acceleration factor of 16, respectively - statistically significant improvements over the original model, which resulted in 0.867\textpm0.025 and 0.949\textpm0.025. Finally, this paper shows that the proposed network not only improves the overall image quality, but also improves the image quality for the regions-of-interest: liver, kidneys, and spleen; as well as generalises better than the baselines in presence of a needle.

\end{abstract}


\end{frontmatter}


\section{Introduction}
\label{sec:introduction}
Medical imaging advances have significantly improved the quality of diagnosis, consequently improving medical care. Non-invasive imaging modalities, such as magnetic resonance imaging (MRI), computed tomography (CT), positron emission tomography (PET), or ultrasonography (US), have proven to be some of the most crucial tools in the arsenal of radiologists~\citep{acharya1995biomedical}. 

MRI is one of the most significant innovations in the field of medical imaging, which uses the principles of nuclear magnetic resonance (NMR) to image hydrogen nuclei~\citep{ahmad2014medical}. Non-ionising radiation and excellent soft-tissue contrast have made MRI one of the most popular imaging modalities over the past few decades. However, due to the sequential nature of the acquisition process, MRI is an inherently slow process - causing it to have less patient throughput and also making it difficult for real-time applications such as interventions. Ignoring parts of the data (known as undersampling) can make the acquisition faster, but consequently, the acquired images might lose resolution or might have imaging artefacts due to the violation of the Nyquist criterion~\citep{lustig2008compressed,chatterjee2021reconresnet}.  

CT is a non-invasive volumetric imaging technique, which shoots a beam of X-rays through the body, which is then captured using a detector present directly opposite to the source to obtain a projection image. Multiple such projections are captured from different angles by rotating the X-ray source and the detector, which are then combined to construct the 3D volume. Unlike MRI, the acquisition speed of CT is fast, making it a commonly used modality for interventional image guidance~\citep{furlow2019ct}. However, CT uses X-rays, which produce harmful ionisation that can cause cancer~\citep{albert2013radiation}. The amount of radiation can be reduced by reducing the current of the X-ray tube or the number of projections, both of which consequently reduce the image quality. 

\subsection{Mathematical Preliminaries}\label{sec:background}
\subsubsection{CT}
As described in the previous section, CT imaging works by reconstructing volumes from a sequence of (usually one- or two-dimensional) X-ray projections. A projection can be described mathematically as the Radon transform:
\begin{align}
    p(t,\mathbf{\theta}) = \int_{\mathbb{R}^n}f(\mathbf{x})\delta(\langle\mathbf{x},\mathbf{\theta}^\perp\rangle-t)\,\mathrm{d}\mathbf{x},
\end{align}
that is, the line integral of the volume of attenuation coefficients $f$ along the line in direction $\mathbf{\theta}\in S^{n-1}\subset\mathbb{R}^n$ and the distance from the origin $t\in\mathbb{R}$. For the sake of simplicity, we will assume $n=2$ here, which allows defining $\mathbf{\theta}$ as vector function $\mathbf{\theta}(s) = (-\sin(s), \cos(s))^T$ and consequently $\mathbf{\theta}^\perp(s)=(\cos(s), \sin(s))^T$. Arranging projections along the angular axis $s$ results in so-called \textit{sinograms}. There are different algorithms to reconstruct the attenuation function $f$ from the measured projections. One of the simplest algorithms for 2D reconstructions is the filtered back-projection (FBP), which can be described mathematically as:
\begin{align}
    \tilde{f}(\mathbf{x}) = \frac{1}{2\pi}\int_0^\pi (p(\cdot, \mathbf{\theta}(s))*h)(\langle\mathbf{x},\mathbf{\theta}^\perp(s)\rangle)\,\mathrm{d}s,
\end{align}
using a one-dimensional filtering function $h$ (usually a ramp filter or modifications) convolved with the projections along the detector axis. One can easily see that $\tilde{f}$ is merely an approximation of the original attenuation function $f$ because the measured projections are inherently discrete such that the integral can only be evaluated at discrete positions. Moreover, the filtering function $h$ is unbounded, whereas $p$ is not. This makes the reconstruction an ill-posed problem and leads to slight or severe artefacts, e.g. streak artefacts, depending on the sampling. The FBP was originally developed to reconstruct slices from parallel-beam projections. However, modern CT systems acquire projections using fan-beam (or cone-beam) projections, making them much faster. The FBP is also capable of reconstructing these kinds of projections with small modifications and is, compared to iterative approaches, very fast.

\subsubsection{MRI}
MR images are not directly acquired in the image space, but in the frequency space, the so-called k-space - which is the 2D or 3D Fourier transform of the measured MR image. The spatial frequency can be defined with the following equations:
\begin{align}
    k_{{\mathrm{FE}}}={\bar{\gamma}}G_{\mathrm{FE}}m\Delta t
\label{eq:kFE}\\
    k_{\mathrm{PE}}={\bar{\gamma }}n\Delta G_{{\mathrm{PE}}}\tau
\label{eq:kPE}
\end{align}
where FE refers to frequency encoding, PE refers to phase encoding, $\bar {\gamma }$ is the gyromagnetic ratio, $G_{{\mathrm {FE}}}$ is the number of gradient steps in the frequency encoding direction, $G_{{\mathrm  {PE}}}$ is the number of gradient steps in the phase encoding direction, $\Delta t$ is the sampling time, $\tau$  is the duration of $G_{\mathrm{PE}}$, $m$ is the number of samples in the FE direction, $n$ is the number of samples in the PE direction. The 2D Fourier transform of the signal represents the spin density distribution in two dimensions, constituting the positional coordinate ($x$, $y$) and spatial frequency ($k_{\mathrm{FE}}$, $k_{\mathrm{PE}}$) a Fourier transform pair~\citep{mcrobbie2017mri}. 

This k-space can be sampled in various ways: Cartesian, radial, or spiral sampling, to name a few. Cartesian sampling is the most frequently used sampling strategy. However, images acquired using this might suffer from motion artefacts if not handled explicitly by the acquisition sequence. Radial sampling has an edge by being inherently robust against motion. In radial imaging, 1D projections (known as spokes) are acquired around the object from different angles to fill the k-space. But unlike CT reconstruction, radially acquired MRIs are reconstructed by ``re-gridding'' them on a conventional Cartesian grid before reconstructing them using the inverse Fourier transform or using a non-uniform fast Fourier transform (NUFFT)~\citep{liu1998accurate,fessler2007nufft}. 

If the k-space is not sampled densely enough (undersampling), i.e. sampled less than the Nyquist criterion, the reconstructed image might contain undersampling artefacts. This is applicable for any sampling pattern, though the type of artefact varies. The artefact that is usually present in an undersampled radial MRI is the streak artefact.

\subsubsection{Fourier Slice Theorem}\label{ssec:fourier_slice}
CT and radial MRI measurements are mathematically very similar. The Fourier slice theorem tells that the one-dimensional Fourier transform ($\mathcal{F}_1$) of a parallel-beam projection for a certain angle ($p_\theta$) is the same as extracting a one-dimensional central slice with the same angle ($S_\theta$) from the two-dimensional Fourier transform ($\mathcal{F}_2$) of the image function ($f$):
\begin{align}
    \mathcal{F}_1[p_\theta](u) = (S_\theta\circ\mathcal{F}_2[f])(u)
\end{align}
This means that every parallel-beam CT projection fills up the 2D Fourier space by one spoke. The spokes acquired in radial MRI do exactly the same which means that conversely, applying the one-dimensional inverse Fourier transform on these spokes results in values similar to CT projections, such that it is possible to use CT reconstruction algorithms like FBP to ultimately reconstruct MRI images. This was the earliest method to reconstruct such MRIs~\citep{mcrobbie2017mri}.

Though not directly applicable to the more commonly used fan- and cone-beam projections, the Fourier slice theorem has been used in reconstruction algorithms for these beam types with slight modifications \cite{Zhao95}.

\subsection{Related Work}\label{sec:related}
There is a variety of approaches for CT reconstruction of sparse projections using deep learning. Post-processing networks aim to reduce artefacts caused by the sparse sampling after reconstructing with a traditional algorithm like FBP, but usually lack data consistency and depend on the quality of the initial reconstruction. Examples include FBPConvNet~\citep{jin2017deep}, ISTA UNet~\citep{Liu2020interpreting} or Framing UNet~\citep{han_framing_2017}. Dual Branch Prior-Net~\citep{Ghosh22} makes use of a high quality planning scan to improve the quality of the sparse view CT reconstruction. Conversely, another type of approach seeks to optimise or inpaint the sinogram \textit{before} applying a reconstruction algorithm, such as the Deep-neural-network based sinogram synthesis~\citep{lee_dnn_2019} or the Band Patch Network~\citep{Okamoto22}, hence ensuring data consistency. Finally, unrolled iterative reconstruction algorithms can be seen as the third type, which reduce the artefacts in the reconstructed image by optimising both sinograms and reconstructions in an alternating manner after projecting and back-projecting in each iteration, respectively. These algorithms include the Primal-Dual network~\citep{adler2018learned}, LEARN~\citep{chen_learn_2018} or Deep Convolutional Gated Recurrent Units~\citep{Ikuta22}. \citet{leuschner2021quantitative} presented a quantitative comparison of these and other related algorithms for low-dose and sparse CT reconstructions. They have reported the Primal-Dual network as the best performing model based on the overall performance.

Various approaches, including deep learning based techniques, have been proposed to reconstruct undersampled radial MRI. UNet is a commonly used network architecture to reconstruct undersampled MRI~\citep{ChangMinHyun.2018}. The NCC1701~\citep{chatterjee2021reconresnet} pipeline, which includes ReconResNet as the backbone, has shown an improvement over UNet while reconstructing undersampled Cartesian and radial MRIs. These approaches work in the image space, i.e. the undersampled images are first reconstructed with adjoint NUFFT, and then the neural network model is applied to remove the artefacts. On the other hand, Automap~\citep{zhu2018image} works by taking the k-space as input and providing the reconstructed image as output. But the network design of Automap creates a hindrance for scalability. Several other deep learning based approaches have been proposed, which deal with undersampled radial reconstruction, working in the image space or k-space. However, the path of reconstructing them by converting the radial k-space to a sinogram, then applying sinogram synthesis or sinogram correction techniques before applying FBP has not been explored in depth.

\subsection{Contributions}\label{sec:contrib}
This paper improves the state-of-the-art Primal-Dual network with the novel Primal-Dual UNet model in terms of accuracy and speed of reconstruction. The proposed method has been evaluated for the task of sparse CT reconstruction for three different sparsity levels: 4, 8, and 16, and has been compared against three deep learning based baseline models. Moreover, this research also shows the potential of using sparse CT reconstruction techniques for reconstructing undersampled radially-acquired MRI by performing spoke-wise 1-dimensional inverse fast Fourier transform as a pre-processing step. The proposed method, as well as three other sparse CT reconstruction techniques (including two deep learning based methods) and a well-established deep learning based MRI reconstruction model, are compared for the task of undersampled MRI reconstruction for acceleration factors 8 and 16 using two different datasets of two different organs, containing two different MRI contrasts. In terms of evaluating the final results, this paper not only evaluates the overall image quality, but also evaluates the image quality for three regions-of-interest: liver, kidneys, and spleen. Finally, evaluations were performed by comparing the performance of the proposed method against the baselines in the presence of a needle - without explicitly training the models with needles. These additional evaluations are performed to demonstrate the suitability of this approach in clinical scenarios - for routine scans, as well as image-guided interventions. 

\section{Methodology}\label{sec:Methodology}

\subsection{Proposed Architecture}
The network architecture proposed in this paper is based on the Primal-Dual Network~\citep{adler2018learned}, which achieves superior reconstruction quality compared to other deep learning based reconstruction algorithms~\citep{leuschner2021quantitative}, and is combined with UNet~\citep{Ronneberger2015}, which is well-known and often used in medical image processing.

The Primal-Dual Network can be interpreted as an unrolled iterative reconstruction algorithm that optimises in both the sinogram and image space using blocks of fully convolutional layers. In each iteration, the processed sinograms are reconstructed using FBP, combined with the processed images, and reprojected to be combined with the previous sinograms. This does not only improve the quality of the final reconstructions, but also ensures data consistency with the original sinogram. As with many iterative algorithms, the quality of the reconstructions of the Primal-Dual Network depends on the number of iterations, i.e. for a fixed number of parameters in an iteration, an optimal (usually minimal) number of iterations has to be found for the network to converge. If the number of parameters in the convolutional blocks is low, more iterations are needed for convergence. This, however, increases the processing time, with the reconstruction and projection operators being the bottleneck. On the other hand, if the number of parameters in the convolutional blocks increases, much fewer iterations are needed for convergence. The processing time is not much different though, since the convolutions are the bottleneck in this case.

The idea of the network architecture proposed in this paper tackles this trade-off by (1) keeping the number of iterations low, i.e. the reconstruction and projection operators are not a bottleneck, and (2) replacing the convolutional blocks in the image space with a UNet, to get a high number of parameters while keeping the processing time low. This architecture will be referred to as \textit{PD-UNet}.

\subsection{Baselines}
\label{sec:baselines}
The performance of the proposed model was compared against three deep learning based methods. First, a UNet~\citep{Ronneberger2015} was applied on reconstructed Undersampled CT images~\citep{jin2017deep} and MRIs~\citep{ChangMinHyun.2018}, referred to hereafter as \textit{Reconstruction UNet}. The CT images were reconstructed using FBP, and the undersampled MRIs were reconstructed using adjoint NUFFT~\citep{Lin.2018}. The undersampled images were supplied as input to the Reconstruction UNet model, and the outputs of the model were compared against the ground-truth fully-sampled images to calculate the loss while training. 

The second baseline was the \textit{Sinogram UNet}~\citep{lee2018deep}. For this method, the sparsely sampled sinograms were upsampled using bilinear interpolation before supplying them to the UNet model as input, and the loss was computed by comparing the outputs of the model against the corresponding fully-sampled sinograms. 

The final deep learning baseline was the \textit{Primal-Dual Network}~\citep{adler2018learned}, where the sparsely sampled sinograms and zero-initialized reconstructions were supplied as input to the model, and the outputs from the model were compared against the ground-truth fully-sampled images to calculate the loss.

Lastly, the performance of the models was compared with FBP reconstructions after upsampling the sparse-sampled sinograms using bilinear interpolation. For MRI, one additional comparison was performed against images obtained by applying adjoint NUFFT~\citep{Lin.2018} on the undersampled radial k-space (referred to in this paper as \textit{Undersampled (NUFFT)}). It is noteworthy that the majority of the baselines (Sinogram UNet, Primal-Dual Network, and Sinogram Bilinear) were never used before for undersampled MR reconstruction.

\begin{figure*}[!htbp]
\centering
\includegraphics[width=0.95\textwidth]{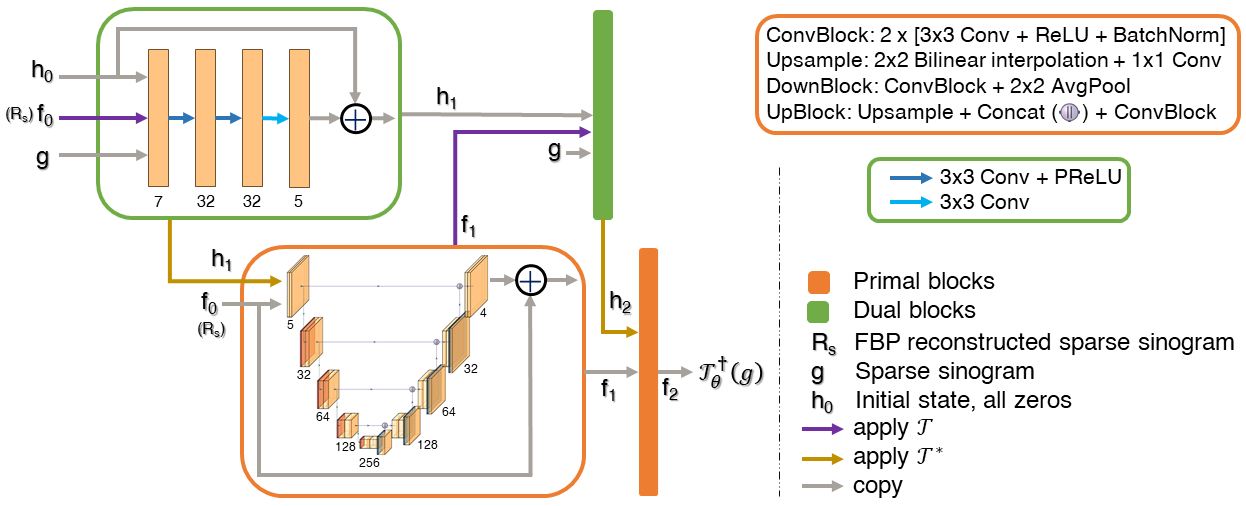}
\caption{Primal-Dual UNet (PD-UNet): Architecture of the proposed network. The orange and the green boxes signify the primal and dual iterates, respectively. The proposed primal block uses a UNet model instead of a fully-convolutional network. In contrast, the dual block is the same fully-convolutional network as the original primal-dual network. It is to be noted that the second orange and green blocks have the same architecture as the first one. The initial dual block receives the image reconstructed by applying FBP on the sparse sinogram as $f_0$ (unlike the original primal-dual network, which receives all zeros), all-zeros as the initial-state $h_0$, and the sparse sinogram $g$. The output of this block, along with the same $f_0$ is given as input to the first primal block.}
\label{netarch}
\end{figure*}

\subsection{Data Normalisation} \label{sec:norm}
Data in the image space was normalised by dividing each slice by the $99^{th}$ percentile of the intensity values present in the total training and validation sets. The $99^{th}$ percentile was chosen to resemble the largest range of the tissue intensities while excluding outliers caused by, e.g., noise or foreign bodies like interventional instruments, and still restricting the input range to roughly the unit interval to enable faster convergence of the network training. Especially in the case of CT data, where intensity values directly correspond to types of tissues, this kind of global normalisation enables the networks to learn specific relations based on the intensity histogram, whereas a per-slice normalisation scheme (e.g. per-slice min-max normalisation) would discard this potentially useful information.

Intensity values of sinograms do not have a trivial relation to tissue intensities. Therefore and due to prior experiments, they were normalised using the Z-score normalisation method - by applying the following equation on each sinogram:
\begin{equation}
\begin{aligned}
    {S_{N}} = \frac{S - \mu_s}{\sigma_s}
\end{aligned}
\label{eq:znorm}
\end{equation}
where $S_N$ is the normalised sinogram, $S$ is the original sinogram, and $\mu_s$ and $\sigma_s$ are the mean and standard deviation of the values present in the sinogram.

For the Reconstruction UNet and Sinogram UNet, the input was normalised using the image space normalisation and sinogram normalisation methods, respectively. The output given by those models were "de-normalised" to obtain the final output - for image space, multiplying with the $99^{th}$ percentile of the intensity values; for sinogram, by applying the following equation:

\begin{equation}
\begin{aligned}
    {S} = {(S_{N} \times \sigma_s) + \mu_s}
\end{aligned}
\label{eq:zunnorm}
\end{equation}

For the Primal-Dual Network and the proposed PD-UNet, several normalisation and de-normalisation steps were performed. Initially, $\mu_s$ and $\sigma_s$ were calculated from the sparsely sampled input sinogram, and all the sinogram normalisations and de-normalisations were performed using these values. Each time before a sinogram or an image was given as input to any block, they were normalised using sinogram and image normalisation methods as discussed earlier, and after receiving output from that block, they were de-normalised. This was performed to preserve the relationship between image and sinogram values while using two different types of normalisation techniques for two different data spaces. Before providing the final output of these models, the values were also de-normalised using the image space technique.

\subsection{Implementation and Training} \label{sec:implement}
The models (the proposed model and the baseline models) were trained for 151 epochs with an effective batch size of 32, the best epoch was chosen based on the validation loss and was used for inference on the test set. The number of epochs was chosen by observing the convergence of the validation loss curves of the models in prior experiments. The memory requirements of the proposed model and the different baseline models are not the same, making it impossible to have the same batch size for the different models. To achieve a constant effective batch size for all the models, instead of the conventional "forward-pass then backward-pass" technique for each training step, multiple forward passes were performed, the gradients were summed up, and finally, an accumulated backward-pass was performed. The number of forward passes to be performed before an accumulated backward pass was calculated as: (32 $\div$ actual batch size of the model). To train the models, the loss was calculated using mean absolute error (L1-Loss), and it was minimised using Adam optimiser (learning rate $= \num{1e-3}$, $\beta_1 = 0.9, \beta_2 = 0.999, \epsilon = \num{1e-08}$). The models were implemented using PyTorch~\citep{NEURIPS2019_9015}, with the help of PyTorch~Lightning~\citep{falcon2019pytorch}; and were trained with mixed precision~\citep{micikevicius2017mixed} using Nvidia RTX 2080Ti and Nvidia RTX A6000 GPUs. The code of this project is available on GitHub\footnote{PD-UNet on GitHub:~\url{https://github.com/phernst/pd-unet}}.

\subsection{Dataset} \label{sec:dataset}
\subsubsection{CT}\label{ssec:ct}
In this study, the data of 28 subjects of the CT Lymph Nodes~\citep{ctlymph} collection from The Cancer Imaging Archive~\citep{Clark2013} was used, consisting of reconstructed volumes of the abdomen that serve as ground truth. 16, 4 and 8 subjects were used for training, validation and test set, respectively. Since the number of axial slices and the voxel sizes differed per subject, the central 200 slices of each subject were extracted and interpolated using sinc interpolation to have an in-plane matrix size of 256x256. The $99^{th}$ percentile of this data set was computed to be an attenuation value corresponding to 270 HU, setting the focus of the reconstruction task to tissues with attenuation values less than bones (including soft tissues, fat, blood and other liquids).

The sinograms were simulated using torch\_radon~\citep{torch_radon} as the datasets only contain reconstructions. Each fan-beam projection consists of 511 detector pixels with a spacing of \SI{1}{px}, a source-to-isocenter distance of \SI{400}{px} and a detector-to-isocenter distance of \SI{150}{px} to cover the full axial slice. The sinograms contain 360 equiangular projections with an angular distance of \ang{1} between consecutive projections. The reason for choosing fan-beam over parallel-beam projections here is its applicability to real-world data. However, results for parallel-beam projections can be found in Appendix~\ref{sec_appendix_parallel}.

\subsubsection{MRI}
\label{sec:mri_dataset}
Two different publicly available benchmark datasets of two different organs were used in this research: IXI dataset\footnote{IXI Dataset: \url{https://brain-development.org/ixi-dataset/}.} for brain and CHAOS challenge dataset~\citep{kavur2021chaos} for abdomen. The IXI dataset contains nearly 600 brain MRIs of normal healthy subjects, acquired using different MRI protocols (T1w, T2w, PDw, MRA, and DWI), collected from three different hospitals at two different field strengths (1.5T and 3T). 30 central slices from 100, 35 and 50 T1w volumes acquired at 3T were used in this study as the training, validation and test set, respectively. The CHAOS challenge dataset contains abdominal MRIs of 40 healthy subjects, acquired using two different MR sequences: T1-Dual (In-phase and Opposed-phase) and T2-SPIR. All the slices from 24, 6 and 10 subjects (three volumes each: T1-in, T1-opposed, and T2) were used in the training, validation, and test set, respectively. All the images were interpolated with sinc interpolation to have an in-plane matrix size of 256x256. 

The datasets do not contain any raw MRI data, only the magnitude images, which were treated as the fully-sampled groundtruth. The corresponding single-coil radial k-spaces of those magnitude images were generated using NUFFT (implemented in PyNUFFT~\citep{Lin.2018}). The fully-sampled raw data was considered to have the number of spokes (radial acquisitions) as twice the base resolution, which was 512 for this dataset. The sampling was performed following the equidistant radial sampling scheme, where the angle between the spokes, calculated as $\Delta\phi=\pi\div{n_{Sp}}$ for $n_{Sp}\in\mathbf{N}$ spokes, was $\Delta\phi=\ang{0.3515625}$. A NUFFT was performed on each slice of the MRI volumes, following $\Delta\phi$, to obtain the equivalent fully-sampled radial k-space.

\paragraph{Pre-processing}
\label{sec:mri_preprocessing}
The fully-sampled sinograms of the respective radial k-spaces were obtained by applying the one-dimensional inverse Fourier transform on each spoke of the k-space. To keep the setup similar to the one described in Sec.~\ref{ssec:ct}, the spokes were shifted by half a detector pixel using sinc interpolation and were cropped to the central 363 detector pixels. This step was necessary because each spoke extracted by NUFFT contained 512 frequency components, which corresponds to a detector pixel number of 512, as well. In contrast to the simulated CT projections in Sec.~\ref{ssec:ct}, the MR sinograms correspond to parallel-beam projections according to the Fourier slice theorem (cf. Sec.~\ref{ssec:fourier_slice}).

\subsubsection{Undersampling}
\label{sec:mri_undersampling}
To simulate the undersampled datasets, the sinograms were made sparse by retaining only every $n^{th}$ projection (spoke), where $n$ denotes the level of sparsity. For CT, three different sparsity levels were experimented with: 4, 8 and 16 (referred herewith as Sparse 4, Sparse 8 and Sparse 16, respectively). In case of MRI, two levels were used: Sparse 8 and Sparse 16.

\subsection{Evaluation criteria} \label{sec:evalcriteria}
The performance of the models was evaluated and compared quantitatively with the help of root-mean-squared error (RMSE) and structural similarity index (SSIM)~\citep{wang2004imageSSIM}. Moreover, the statistical significance of the improvements observed was evaluated by the Mann–Whitney U test~\citep{mann1947test}. Finally, they were also compared qualitatively for selective slices with the help of difference images and SSIM maps. Slices that resulted in SSIM values identical to the median value in up to three decimal points for Primal-Dual UNet (the proposed method) and Primal-Dual Network (main baseline) were chosen for qualitative portrayal for each experiment - to be able to choose slices which are representative of the results for each of the experiments.


\section{Results}\label{sec:Results}
  
\subsection{CT}
The performance of the proposed Primal-Dual UNet was compared (see Sec.~\ref{sec:baselines}) both quantitatively and qualitatively against three other deep learning models: Reconstruction~UNet~\citep{jin2017deep}, Sinogram~UNet~\citep{lee2018deep}, and Learned~Primal-Dual~Network~\citep{adler2018learned}, and also against reconstructions with the standard FBP applied on the sinograms up-sampled using bilinear interpolation, referred to here as \textit{Sinogram Bilinear}. The fan-beam geometry being more realistic than the parallel-beam geometry, the focus of this research was on the fan-beam geometry - hence these results are shown in this section. Additional experiments were also performed with the parallel-beam geometry, and the results have been reported in Appendix~\ref{sec_appendix_parallel}. Experiments were performed for three different levels of sparsity: 4, 8, and 16 (referred to as: Sparse 4, 8, and 16, respectively). 

\begin{figure}[h!]
    \centering
    \includegraphics[width=0.48\textwidth]{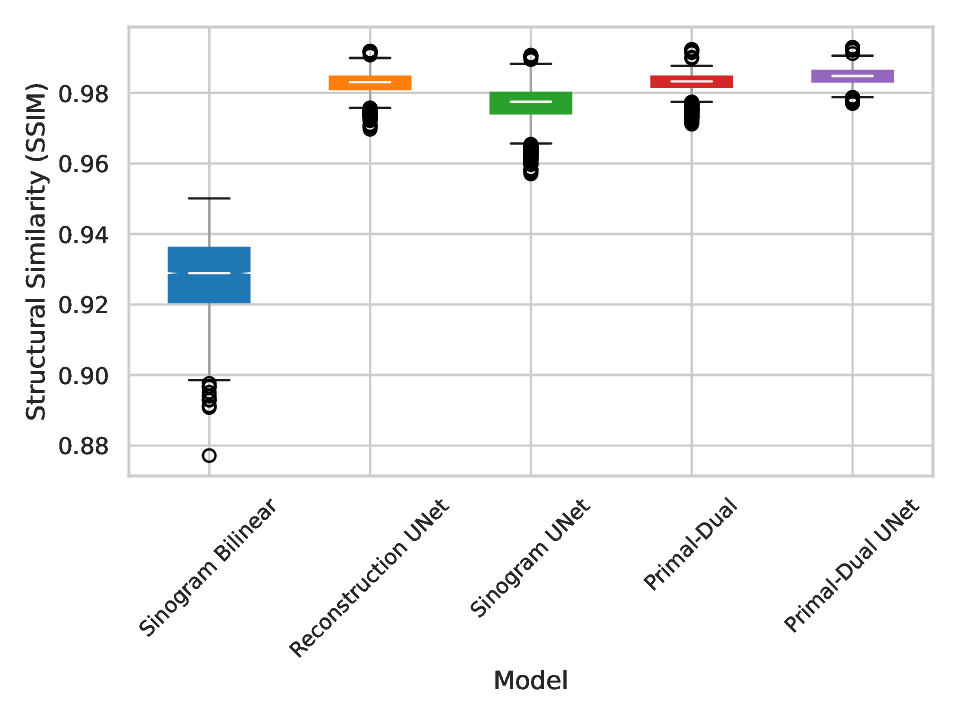}
    \caption{Box-plots of the resultant SSIM values for CT (fan-beam geometry) Sparse 4}
    \label{fig:box_ct_fan_Sparse4}
\end{figure}

\begin{figure}[h!]
    \centering
    \includegraphics[width=0.48\textwidth]{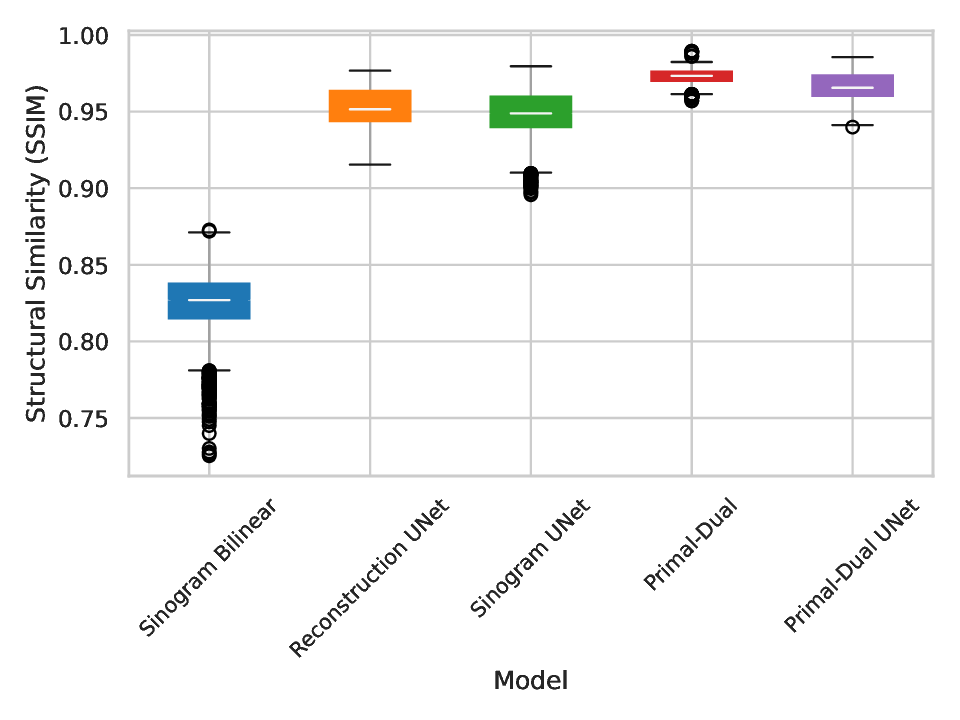}
    \caption{Box-plots of the resultant SSIM values for CT (fan-beam geometry) Sparse 8}
    \label{fig:box_ct_fan_Sparse8}
\end{figure}

\begin{figure}[h!]
    \centering
    \includegraphics[width=0.48\textwidth]{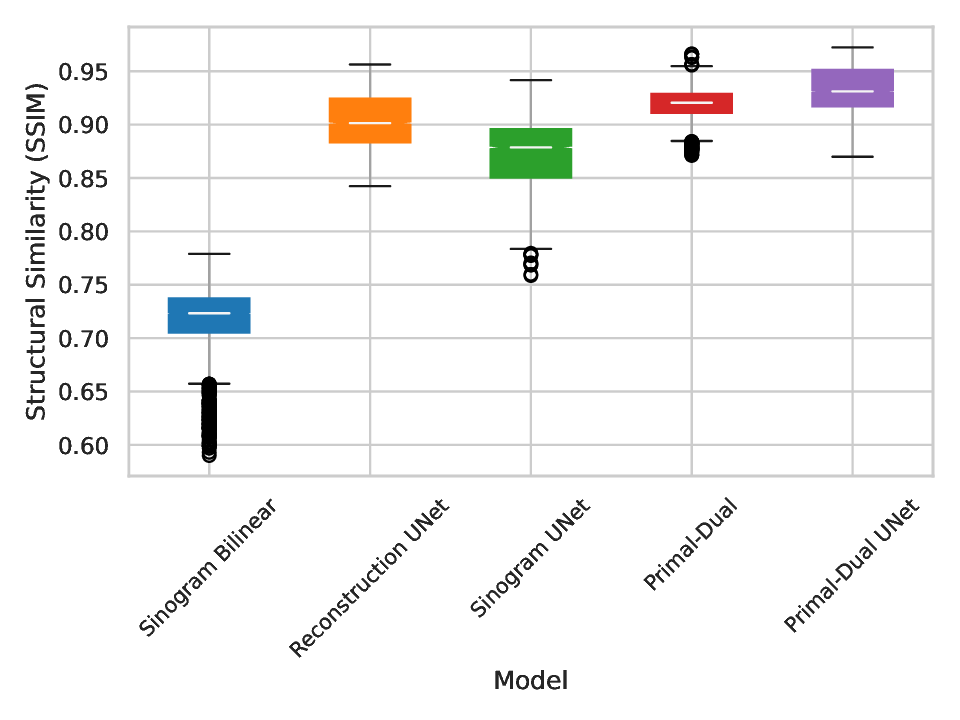}
    \caption{Box-plots of the resultant SSIM values for CT (fan-beam geometry) Sparse 16}
    \label{fig:box_ct_fan_Sparse16}
\end{figure}

Quantitative evaluations were performed using RMSE, calculated in the Hounsfield scale, and SSIM, calculated on the normalised intensity values, as shown in Tab.~\ref{tab:res_CT_fan}. The range of the resultant SSIM values is portrayed with the help of box plots in Fig.~\ref{fig:box_ct_fan_Sparse4}, \ref{fig:box_ct_fan_Sparse8}, and \ref{fig:box_ct_fan_Sparse16} for the three different levels of sparsity.

\begin{table*}[!htbp]
\centering
\caption{Resultant metrics for CT fan-beam geometry (mean\textpm std)}
\label{tab:res_CT_fan}
\begin{tabular}{@{}lcccccc@{}}
\toprule
\multicolumn{1}{c}{\multirow{2}{*}{Method}}                           &
  \multicolumn{3}{c}{SSIM} &
  \multicolumn{3}{c}{RMSE (Hounsfield units, HU)} \\ \cmidrule(r){2-4}\cmidrule(l){5-7} 
&
  \multicolumn{1}{c}{Sparse 4} &
  \multicolumn{1}{c}{Sparse 8} &
  Sparse 16 &
  \multicolumn{1}{c}{Sparse 4} &
  \multicolumn{1}{c}{Sparse 8} &
  Sparse 16 \\ \midrule
Bilinear Sinogram &
  \multicolumn{1}{c}{0.928\textpm0.011} &
  \multicolumn{1}{c}{0.824\textpm0.021} &
  0.716\textpm0.033 &
  \multicolumn{1}{c}{33.135\textpm4.557} &
  \multicolumn{1}{c}{59.588\textpm7.645} &
  90.148\textpm12.033 \\
Sinogram UNet &
  \multicolumn{1}{c}{0.977\textpm0.005} &
  \multicolumn{1}{c}{0.948\textpm0.016} &
  0.874\textpm0.032 &
  \multicolumn{1}{c}{14.482\textpm2.324} &
  \multicolumn{1}{c}{26.811\textpm13.810} &
  47.574\textpm11.621 \\
Reconstruction UNet &
  \multicolumn{1}{c}{0.983\textpm0.003} &
  \multicolumn{1}{c}{0.953\textpm0.012} &
  0.903\textpm0.026 &
  \multicolumn{1}{c}{\textbf{11.860\textpm1.891}} &
  \multicolumn{1}{c}{25.575\textpm5.384} &
  47.689\textpm12.170 \\
Primal-Dual Network &
  \multicolumn{1}{c}{0.983\textpm0.003} &
  \multicolumn{1}{c}{\textbf{0.973\textpm0.005}} &
  0.919\textpm0.016 &
  \multicolumn{1}{c}{21.693\textpm3.216} &
  \multicolumn{1}{c}{23.868\textpm3.806} &
  35.386\textpm6.212 \\
\textbf{Primal-Dual UNet} &
  \multicolumn{1}{c}{\textbf{0.985\textpm0.002}} &
  \multicolumn{1}{c}{0.966\textpm0.008} &
  \textbf{0.932\textpm0.021} &
  \multicolumn{1}{c}{15.835\textpm2.143} &
  \multicolumn{1}{c}{\textbf{22.343\textpm4.367}} &
  \textbf{34.383\textpm8.788} \\ \bottomrule
\end{tabular}%
\end{table*}

In terms of SSIM, the proposed method achieved improvements over all the baseline methods with statistical significance for Sparse 4 and 16 - including improvements of 0.2\% and 1.39\% respectively over the baseline Primal-Dual Network. However, Primal-Dual Network scored 0.72\% better average SSIM than the proposed method with a statistical significance for Sparse 8. On the other hand, the proposed method scored better RMSEs (27\%, 6.39\%, 2.83\% for Sparse 4, 8, 16) than the baseline Primal-Dual Network with statistical significance for all three levels of sparsities. For Sparse 8 and 16, the proposed method achieved better RMSEs than all the baselines. However, for Sparse 4, both Sinogram UNet and Reconstruction UNet achieved better RMSEs than the baseline Primal-Dual Network, as well as the proposed method.

\begin{figure}[!htbp]
    \centering
    \includegraphics[width=0.48\textwidth]{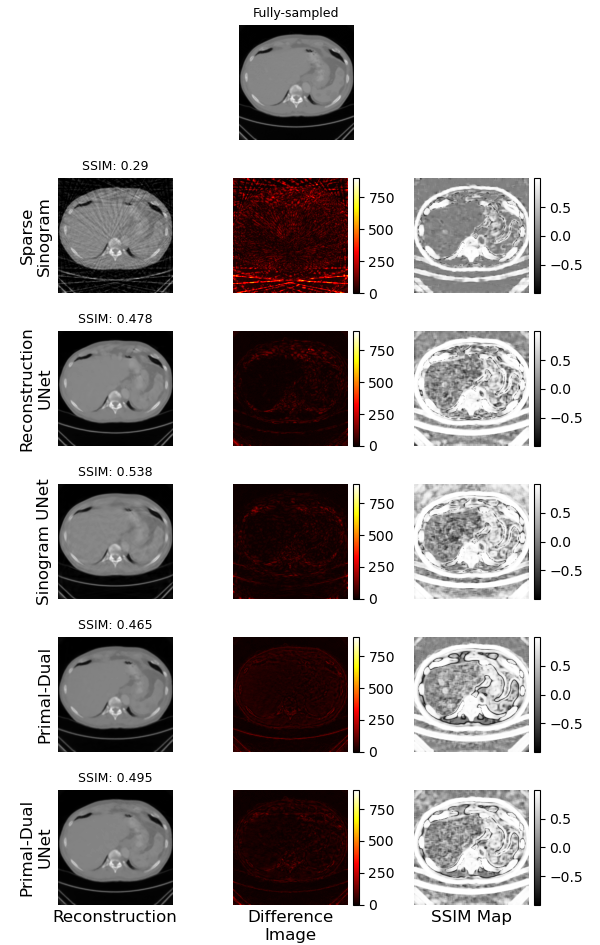}
    \caption{Qualitative comparisons of the reconstructions of CT (fan-beam geometry) Sparse 8}
    \label{fig:res_ct_fan_Sparse8}
\end{figure}

\begin{figure}[!htbp]
    \centering
    \includegraphics[width=0.48\textwidth]{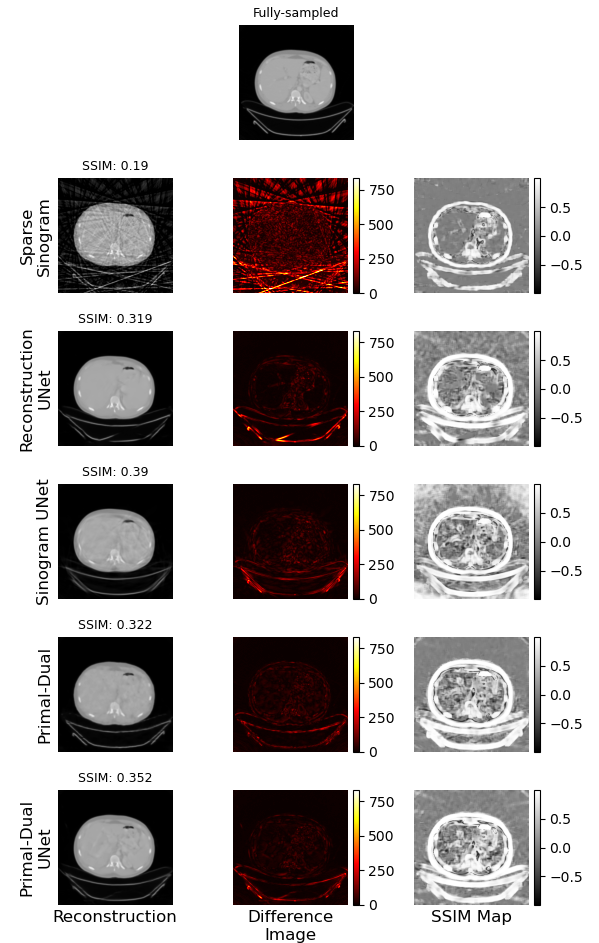}
    \caption{Qualitative comparisons of the reconstructions of CT (fan-beam geometry) Sparse 16}
    \label{fig:res_ct_fan_Sparse16}
\end{figure}

Fig.~\ref{fig:res_ct_fan_Sparse8}, and \ref{fig:res_ct_fan_Sparse16} show qualitative comparisons of the reconstructions for Sparse 8 and 16, respectively, while the Sparse 4 comparison can be found in Appendix (Sec.~\ref{sec_appendix_fan_4}, Fig.~\ref{fig:res_ct_fan_Sparse4}). Comparisons are performed with the help of difference images (in the Hounsfield scale) and SSIM maps (calculated on the normalised intensity values). By looking at the qualitative results, it can be said that they do corroborate with the quantitative results.


\subsubsection{Simulated needle insertion}
\begin{figure}[!htbp]
    \centering
    \includegraphics[width=0.48\textwidth]{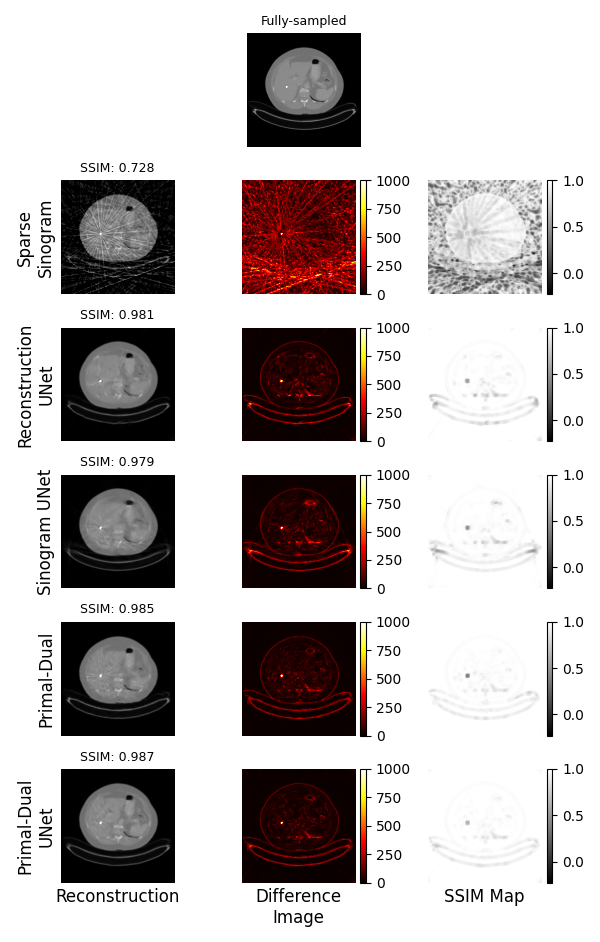}
    \caption{Qualitative comparisons of the reconstructions of CT (fan-beam geometry) Sparse 16 with needle for an exemplary slice.}
    \label{fig:res_ct_fan_Sparse8_needle}
\end{figure}

As a further test on the practical use of the proposed method, the insertion of an interventional needle into the abdominal scans is simulated and evaluated visually and quantitatively. For this purpose, a NeuWave Medical\footnote{Ablation needle: \url{http://www.neuwavemedical.com}} ablation needle was inserted into an abdominal phantom and was scanned with a KIDS-CT scanner. The needle was segmented out of the resulting volume by a simple thresholding. The needle was combined with the available test volumes by summing the attenuation coefficients, which represents a good estimation of actual needle insertion since the values of human tissues and needle materials are significantly different. However, this simulation is missing some artefacts caused by, e.g., photon starvation. These combined volumes served as the ground truth for this experiment. Sparse fan-beam sinograms and reconstructions were again simulated using pytorch\_radon, and the same pre-processing was performed as before. Qualitative and quantitative results of an exemplary slice for Sparse 16 are shown in Fig.~\ref{fig:res_ct_fan_Sparse8_needle}. Despite not being trained on datasets with needles, the networks seem to be capable of reconstructing these highly absorbing materials instead of assuming them to be artefacts to be removed or replaced by soft tissue attenuation coefficients. Similar to the results obtained in the previous experiments without the needle, all networks improve the FBP reconstruction of the sparse sinogram by at least 25 percentage points SSIM. Sinogram UNet still performs worst, followed by Reconstruction UNet, Primal-Dual Network, and the best performing Primal-Dual UNet with 0.979, 0.981, 0.985, and 0.987, respectively.

It is of special interest to evaluate the reconstruction in the region around the needle. For this reason, a 32x32 patch around the needle was extracted from every prediction of all test volumes, and the errors were calculated on this region of interest only. These results can be found in Tab.~\ref{tab:res_CT_fan_needle}.

\begin{table}[!htbp]
\centering
\caption{Resultant metrics for CT fan-beam geometry with inserted needle for Sparse 16 (mean\textpm std)}
\label{tab:res_CT_fan_needle}
\begin{tabular}{@{}lcc@{}}
\toprule
  \multicolumn{1}{c}{Method}&
  SSIM &
  RMSE (HU)\\\midrule
Sparse Sinogram & 0.817\textpm0.015 & 355\textpm8 \\
Bilinear Sinogram & 0.871\textpm0.006 & 523\textpm2 \\
Sinogram UNet & 0.915\textpm0.008 & 430\textpm9 \\
Reconstruction UNet & 0.919\textpm0.011 & 403\textpm3 \\
Primal-Dual Network & 0.913\textpm0.005 & 458\textpm6 \\
\textbf{Primal-Dual UNet} & \textbf{0.940\textpm0.004} & \textbf{367\textpm8} \\ \bottomrule
\end{tabular}%
\end{table}

The trend continues as described previously: a simple upsampling of the sinogram results in a small increase of 5 percentage points SSIM wrt. the FBP of the sparse sinogram. Interestingly, Reconstruction UNet performs slightly better in terms of the average SSIM than Primal-Dual Network, though both still perform reasonably well with more than 0.91 SSIM. The proposed Primal-Dual UNet increases the SSIM of Reconstruction UNet by a large margin, to an average SSIM of 0.940 - an improvement of more than 2 percentage points. This shows that the presented model is not only capable of reconstructing images with higher quality compared to competing reconstruction networks trained on the same dataset but is also likely to have a higher degree of generalisation regarding different kinds of CT datasets and learns better how to reconstruct artefact-bearing CT data in general.

\subsection{MRI}
The proposed Primal-Dual UNet was compared (see Sec.~\ref{sec:baselines}) against the undersampled radial k-space reconstruction using PyNUFFT~\citep{Lin.2018} (referred to as \textit{Undersampled (NUFFT)}) by applying FBP on the corresponding sinograms (obtained by applying 1D-iFFT on each spoke, explained in Sec.~\ref{sec:mri_preprocessing}), referred to here as \textit{Sinogram Bilinear}, and finally, against three deep learning baseline models: Reconstruction~UNet~\citep{ChangMinHyun.2018}, Sinogram~UNet~\citep{lee2018deep}, and Learned~Primal-Dual~Network~\citep{adler2018learned} - for two different publicly available benchmark datasets for two different organs: IXI for T1w brain MRIs and CHAOS for T1-Dual and T2w abdominal MRIs. Experiments were performed for two different levels of undersamplings: with an acceleration factor of 8 and 16 - which, in terms of the sparsity of the corresponding sinograms for equidistant radial samplings (see Sec. \ref{sec:mri_dataset} and \ref{sec:mri_undersampling}), are referred to here as Sparse 8 and 16 - to have the same terminology for both CT and MRI.

\subsubsection{IXI Dataset}
\begin{figure}[h!]
    \centering
    \includegraphics[width=0.48\textwidth]{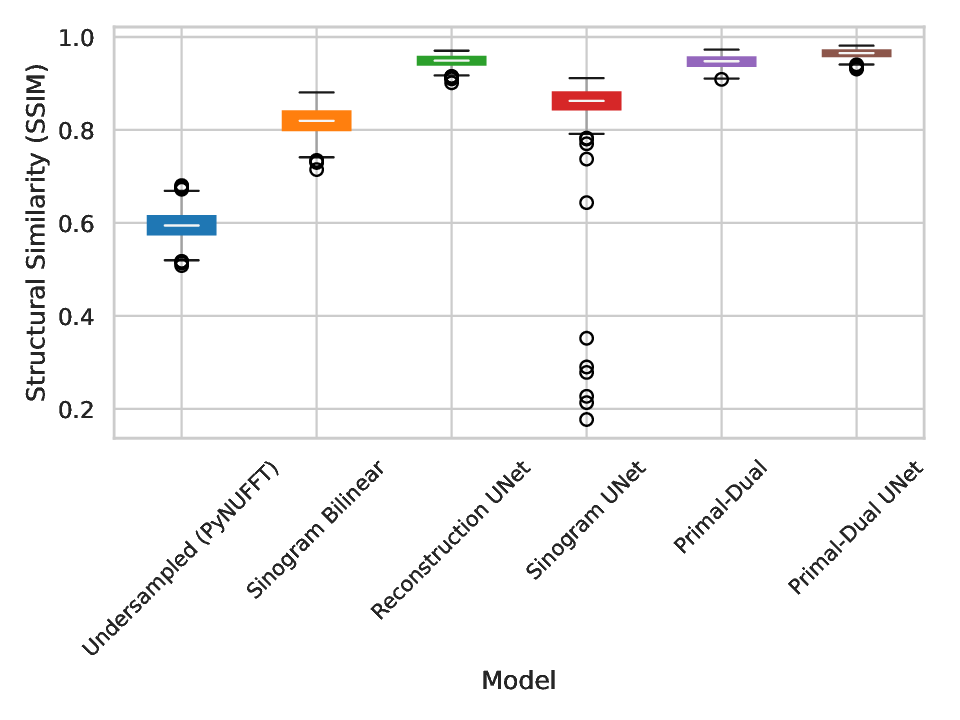}
    \caption{Box-plots of the resultant SSIM values for MRI Sparse 8 for the  IXI dataset}
    \label{fig:box_ixiSparse8}
\end{figure}

\begin{figure}[h!]
    \centering
    \includegraphics[width=0.48\textwidth]{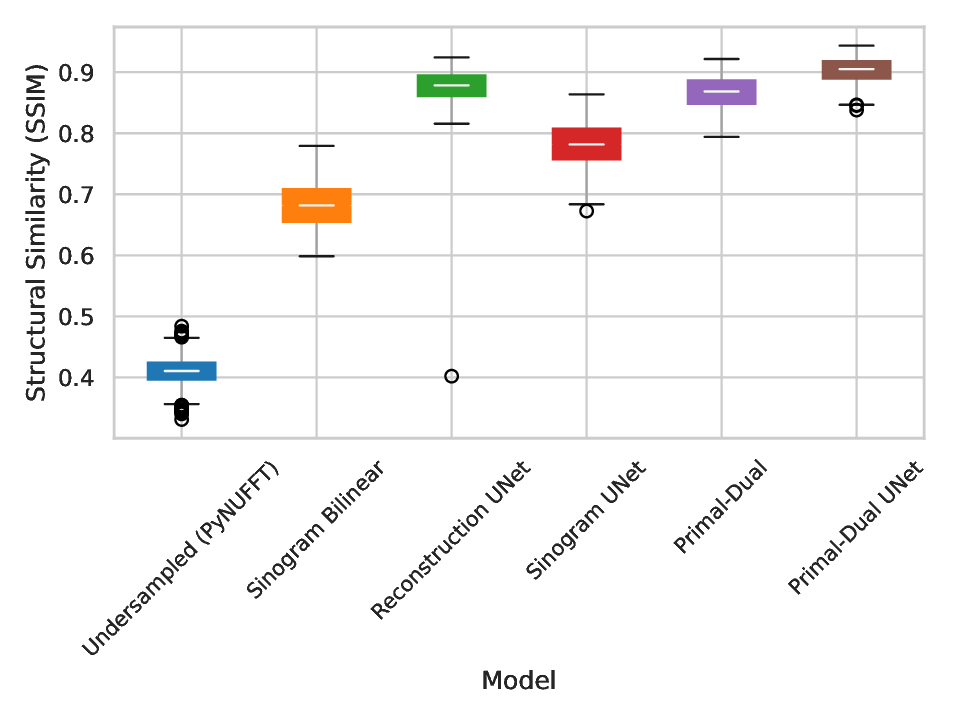}
    \caption{Box-plots of the resultant SSIM values for MRI Sparse 16 for the IXI dataset}
    \label{fig:box_ixiSparse16}
\end{figure}

Qualitative results of SSIM and RMSE are shown in Tab.~\ref{tab:res_IXI}, whereas the range of the resultant SSIM values with the help of box plots is shown in Fig.~\ref{fig:box_ixiSparse8} and \ref{fig:box_ixiSparse16} for Sparse 8 and 16, respectively. It can be observed that the proposed model outperformed all the baseline methods in terms of both SSIM and RMSE, and the statistical tests revealed that these improvements were significant. In terms of average SSIM values, the Primal-Dual UNet achieved improvements of 1.9\% and 4.15\% over the main baseline Primal-Dual Network for Sparse 8 and 16, respectively. Qualitative comparisons of the results using difference images and SSIM maps are shown in Fig.~\ref{fig:res_ixiSparse16} for Sparse 16 and in Appendix (Sec.~\ref{sec_appendix_mri}, Fig.~\ref{fig:res_ixiSparse8}) for Sparse 8.

\begin{table}[h!]
\centering
\caption{Resultant metrics for MRI for the IXI dataset (mean\textpm std)}
\label{tab:res_IXI}
\resizebox{0.48\textwidth}{!}{%
\begin{tabular}{@{}lcccc@{}}
\toprule
\multicolumn{1}{c}{\multirow{2}{*}{Method}}                           & \multicolumn{2}{c}{SSIM}                                        & \multicolumn{2}{c}{RMSE}                                         \\ \cmidrule(r){2-3} \cmidrule(l){4-5}
  & \multicolumn{1}{c}{Sparse 8}             & Sparse 16            & \multicolumn{1}{c}{Sparse 8}             & Sparse 16            \\ \midrule
Undersampled (NUFFT)      & \multicolumn{1}{c}{0.595\textpm0.027}          & 0.410\textpm0.021          & \multicolumn{1}{c}{0.046\textpm0.009}          & 0.085\textpm0.016          \\
Bilinear Sinogram   & \multicolumn{1}{c}{0.819\textpm0.026}          & 0.682\textpm0.035          & \multicolumn{1}{c}{0.058\textpm0.017}          & 0.067\textpm0.016          \\
Sinogram UNet              & \multicolumn{1}{c}{0.860\textpm0.044}          & 0.782\textpm0.032          & \multicolumn{1}{c}{0.058\textpm0.021}          & 0.073\textpm0.023          \\
Reconstruction UNet        & \multicolumn{1}{c}{0.948\textpm0.011}          & 0.877\textpm0.025          & \multicolumn{1}{c}{0.021\textpm0.006}          & 0.037\textpm0.012          \\
Primal-Dual Network & \multicolumn{1}{c}{0.947\textpm0.012}          & 0.867\textpm0.025          & \multicolumn{1}{c}{0.021\textpm0.006}          & 0.041\textpm0.015          \\
\textbf{Primal-Dual UNet}  & \multicolumn{1}{c}{\textbf{0.965\textpm0.008}} & \textbf{0.903\textpm0.019} & \multicolumn{1}{c}{\textbf{0.017\textpm0.005}} & \textbf{0.034\textpm0.011} \\ \bottomrule
\end{tabular}%
}
\end{table}

\begin{figure}[!htbp]
    \centering
    \includegraphics[width=0.48\textwidth]{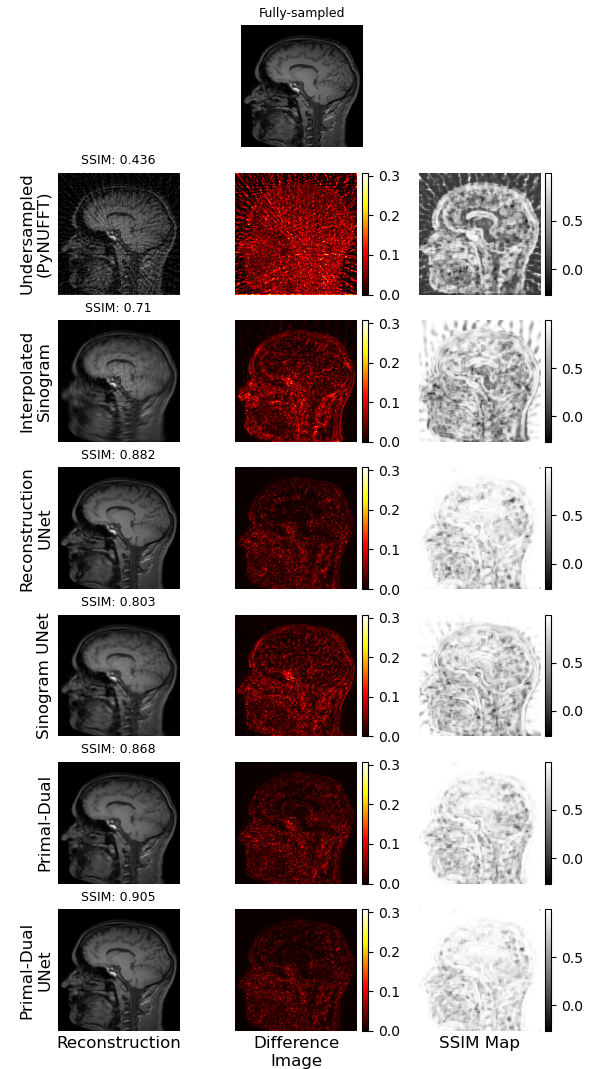}
    \caption{Qualitative comparisons of the reconstructions of MRI Sparse 16 for the IXI dataset}
    \label{fig:res_ixiSparse16}
\end{figure}


\subsubsection{CHAOS Dataset}
\begin{figure}[h!]
    \centering
    \includegraphics[width=0.48\textwidth]{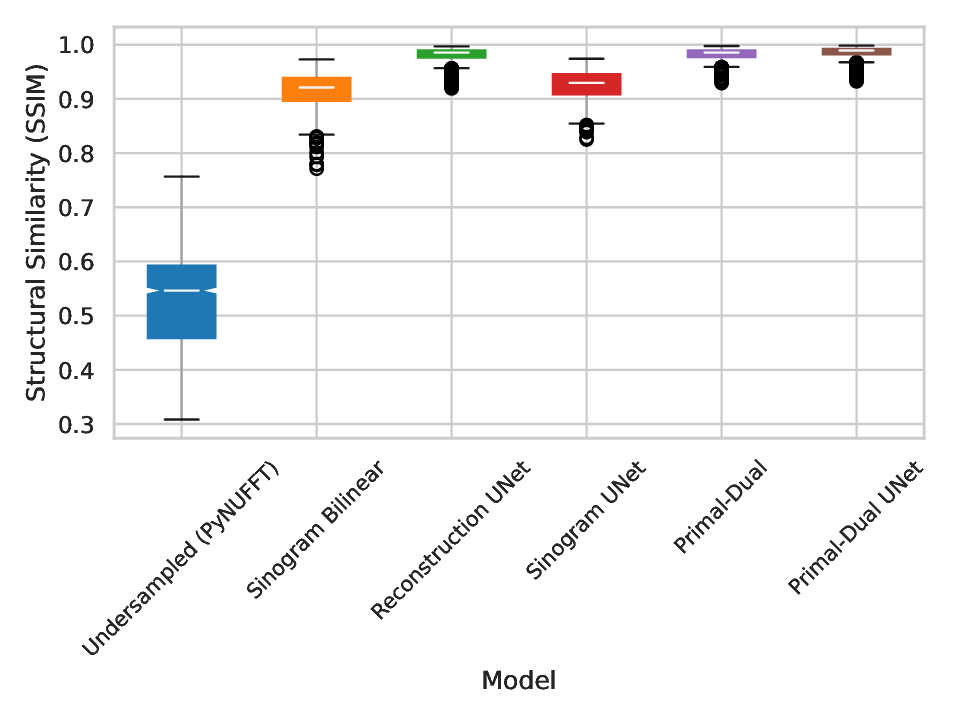}
    \caption{Box-plots of the resultant SSIM values for MRI Sparse 8 for the CHAOS dataset}
    \label{fig:box_chaosSparse8}
\end{figure}

\begin{figure}[h!]
    \centering
    \includegraphics[width=0.48\textwidth]{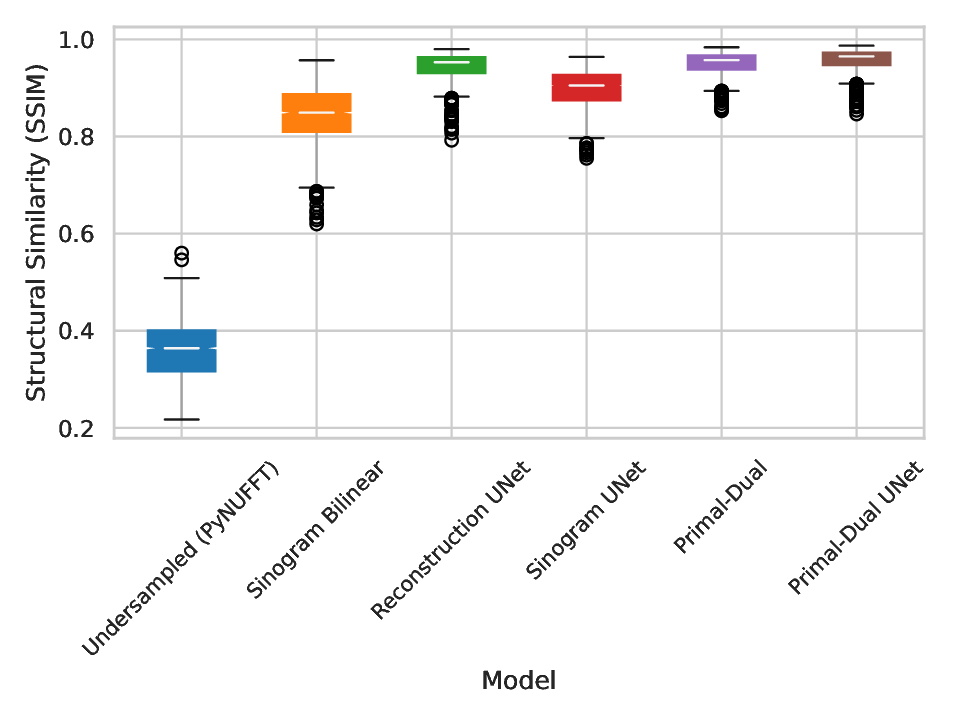}
    \caption{Box-plots of the resultant SSIM values for MRI Sparse 16 for the CHAOS dataset}
    \label{fig:box_chaosSparse16}
\end{figure}

The final set of experiments was performed on the CHAOS MRI dataset. Quantitatively, the proposed method outperformed all the baselines for both acceleration factors with statistical significance when being compared using SSIM and RMSE, reported in Tab.~\ref{tab:res_CHAOS}. Fig.~\ref{fig:box_chaosSparse8} and \ref{fig:box_chaosSparse16} portray the range of resultant SSIM values for Sparse 8 and 16, respectively. The proposed method improved the average SSIM values by 0.41\% and 0.84\% over the baseline Primal-Dual Network, respectively, for Sparse 8 and 16. 


\begin{table}[h!]
\centering
\caption{Resultant metrics for MRI for the CHAOS dataset (mean\textpm std)}
\label{tab:res_CHAOS}
\resizebox{0.48\textwidth}{!}{%
\begin{tabular}{@{}lcccc@{}}\toprule
\multicolumn{1}{c}{\multirow{2}{*}{Method}}                           & \multicolumn{2}{c}{SSIM}
& \multicolumn{2}{c}{RMSE}                                         \\ \cmidrule(r){2-3} \cmidrule(l){4-5}
  & \multicolumn{1}{c}{Sparse 8}             & Sparse 16            & \multicolumn{1}{c}{Sparse 8}             & Sparse 16            \\ \midrule
Undersampled (NUFFT)      & \multicolumn{1}{c}{0.528\textpm0.092}          & 0.359\textpm0.060           & \multicolumn{1}{c}{0.034\textpm0.015}          & 0.068\textpm0.027          \\
Bilinear Sinogram   & \multicolumn{1}{c}{0.915\textpm0.033}          & 0.843\textpm0.060           & \multicolumn{1}{c}{0.035\textpm0.013}          & 0.044\textpm0.016          \\
Sinogram UNet              & \multicolumn{1}{c}{0.926\textpm0.026}          & 0.897\textpm0.038          & \multicolumn{1}{c}{0.032\textpm0.011}          & 0.036\textpm0.014          \\
Reconstruction UNet        & \multicolumn{1}{c}{0.981\textpm0.013}          & 0.943\textpm0.029          & \multicolumn{1}{c}{0.011\textpm0.007}          & 0.022\textpm0.014          \\
Primal-Dual Network & \multicolumn{1}{c}{0.982\textpm0.012}          & 0.949\textpm0.025          & \multicolumn{1}{c}{0.010\textpm0.005}          & 0.021\textpm0.010          \\
\textbf{Primal-Dual UNet}  & \multicolumn{1}{c}{\textbf{0.986\textpm0.011}} & \textbf{0.957\textpm0.023} & \multicolumn{1}{c}{\textbf{0.009\textpm0.006}} & \textbf{0.018\textpm0.011} \\ \bottomrule
\end{tabular}%
}
\end{table}

\begin{table*}[!htbp]
\centering
\caption{Resultant SSIM metrics for different types of MR acquisitions from the CHAOS dataset (mean\textpm std)}
\label{tab:res_CHAOS_subgroup}
\resizebox{\textwidth}{!}{%
\begin{tabular}{@{}lcllcllcc@{}}
\toprule
\multicolumn{1}{c}{\multirow{2}{*}{Method}}                           &
  \multicolumn{4}{c}{Sparse 8} &
  \multicolumn{4}{c}{Sparse 16} \\ \cmidrule(r){2-5}\cmidrule(l){6-9}
 &
  \multicolumn{1}{c}{T2-SPIR} &
  \multicolumn{1}{c}{T1-Dual} &
  \multicolumn{1}{c}{T1 In-phase} &
  T1 Opposed-phase &
  \multicolumn{1}{c}{T2-SPIR} &
  \multicolumn{1}{c}{T1-Dual} &
  \multicolumn{1}{c}{T1 In-phase} &
  T1 Opposed-phase \\ \midrule
Undersampled (NUFFT) &
  \multicolumn{1}{c}{0.599\textpm0.060} &
  \multicolumn{1}{l}{0.493\textpm0.084} &
  \multicolumn{1}{l}{0.430\textpm0.058} &
  0.557\textpm0.053 &
  \multicolumn{1}{l}{0.394\textpm0.051} &
  \multicolumn{1}{l}{0.342\textpm0.057} &
  \multicolumn{1}{c}{0.306\textpm0.043} &
  0.378\textpm0.045 \\
Bilinear Sinogram &
  \multicolumn{1}{c}{0.944\textpm0.017} &
  \multicolumn{1}{l}{0.901\textpm0.029} &
  \multicolumn{1}{l}{0.887\textpm0.029} &
  0.916\textpm0.021 &
  \multicolumn{1}{l}{0.896\textpm0.035} &
  \multicolumn{1}{l}{0.817\textpm0.052} &
  \multicolumn{1}{c}{0.796\textpm0.051} &
  0.838\textpm0.044 \\
Sinogram UNet &
  \multicolumn{1}{c}{0.950\textpm0.014} &
  \multicolumn{1}{l}{0.915\textpm0.023} &
  \multicolumn{1}{l}{0.904\textpm0.023} &
  0.925\textpm0.017 &
  \multicolumn{1}{l}{0.927\textpm0.021} &
  \multicolumn{1}{l}{0.882\textpm0.036} &
  \multicolumn{1}{c}{0.867\textpm0.036} &
  0.898\textpm0.027 \\
Reconstruction UNet &
  \multicolumn{1}{c}{0.983\textpm0.009} &
  \multicolumn{1}{l}{0.980\textpm0.015} &
  \multicolumn{1}{l}{0.975\textpm0.014} &
  0.985\textpm0.013 &
  \multicolumn{1}{l}{0.950\textpm0.019} &
  \multicolumn{1}{l}{0.939\textpm0.033} &
  \multicolumn{1}{c}{0.926\textpm0.031} &
  0.952\textpm0.029 \\
Primal-Dual Network &
  \multicolumn{1}{c}{0.983\textpm0.008} &
  \multicolumn{1}{l}{0.981\textpm0.013} &
  \multicolumn{1}{l}{0.976\textpm0.013} &
  0.986\textpm0.011 &
  \multicolumn{1}{l}{0.955\textpm0.017} &
  \multicolumn{1}{l}{0.945\textpm0.028} &
  \multicolumn{1}{c}{0.934\textpm0.028} &
  0.957\textpm0.022 \\
\textbf{Primal-Dual UNet} &
  \multicolumn{1}{c}{\textbf{0.987\textpm0.007}} &
  \multicolumn{1}{l}{\textbf{0.985\textpm0.012}} &
  \multicolumn{1}{l}{\textbf{0.981\textpm0.011}} &
  \textbf{0.989\textpm0.011} &
  \multicolumn{1}{l}{\textbf{0.961\textpm0.015}} &
  \multicolumn{1}{l}{\textbf{0.955\textpm0.026}} &
  \multicolumn{1}{c}{\textbf{0.945\textpm0.026}} &
  \textbf{0.964\textpm0.023} \\ \bottomrule
\end{tabular}%
}
\end{table*}


The CHAOS dataset includes three different types of MRIs: T1 in-phase, T1 opposed-phase, and T2, acquired using two different sequences for two different contrasts: T1-Dual and T2-SPIR (explained in \ref{sec:mri_dataset}) and all three were combined during training. The metrics that have been shown so far also included all these three types of MRIs together. However, in addition, they were also evaluated separately. Tab.~\ref{tab:res_CHAOS_subgroup} shows the resultant SSIM values for each of the three types of MRIs separately, as well as for both types of acquisition sequences T1-Dual (all the results of T1 in-phase and opposed-phase) and T2-SPIR. The proposed method achieved statistically significant improvements over all the baselines in every scenario. Among the different types of MRIs, T1 opposed-phase achieved the best score, while T1 in-phase achieved the worst. When only the contrasts/sequences are taken into account, then T2-SPIR performed better than T1-Dual. These observations hold true for both levels of sparsity. However, in terms of the percentage of improvement of average SSIM values achieved by the proposed method against the primary baseline model (Primal-Dual Network), the amounts are different for the different levels of sparsity. Both T2-SPIR and T1-Dual improved 0.41\% for Sparse 8, and for Sparse 16 obtained improvements of 0.63\% and 1.06\%, respectively. Considering the three types of MRIs separately, it is noteworthy that even though the T1 in-phase was the worst-performing, it managed to get the most amount of improvement 0.51\% and 1.18\% for Sparse 8 and 16. In contrast, the best-performing type of MRI secured the least amount of improvements: 0.30\% and 0.73\% for Sparse 8 and 16.

\begin{figure}[!htbp]
    \centering
    \includegraphics[width=0.48\textwidth]{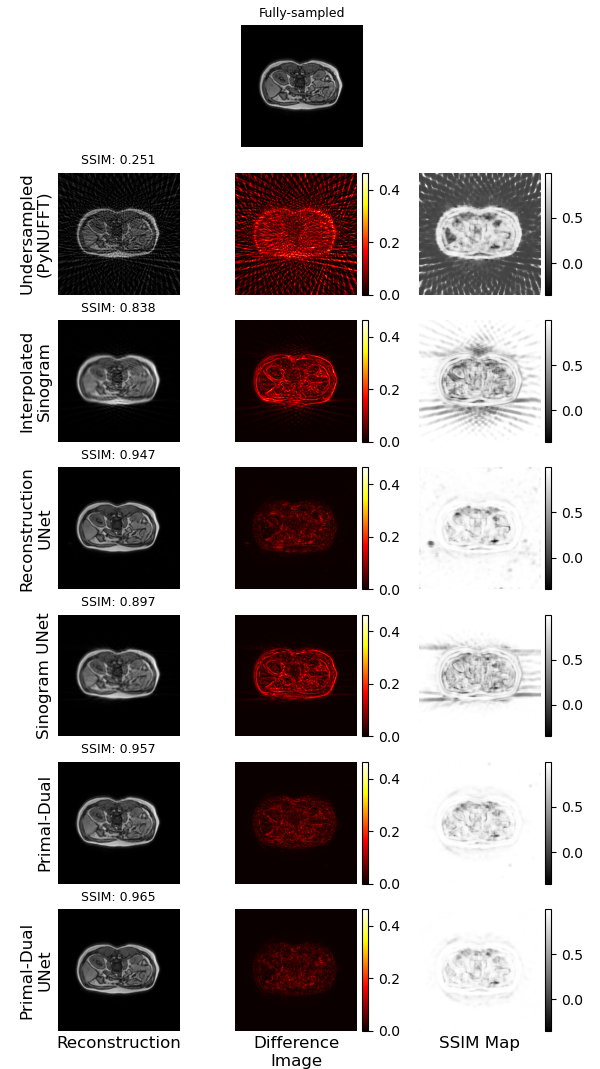}
    \caption{Qualitative comparisons of the reconstructions of CHAOS T1w In-phase Sparse 16 (mean\textpm std)}
    \label{fig:res_chaosT1inSparse16}
\end{figure}

\begin{figure}[!htbp]
    \centering
    \includegraphics[width=0.48\textwidth]{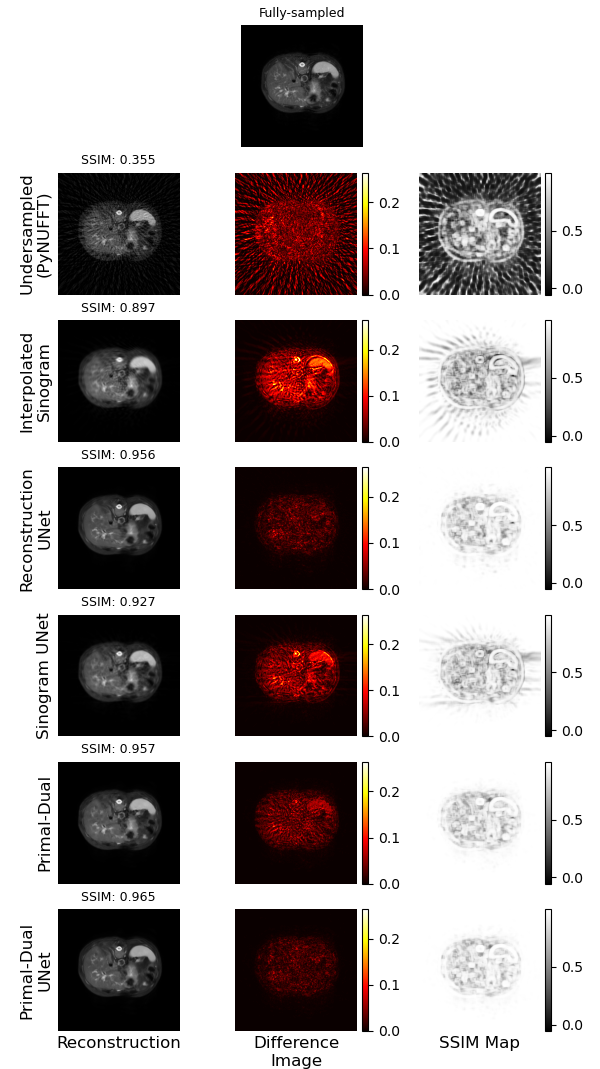}
    \caption{Qualitative comparisons of the reconstructions of CHAOS T2w Sparse 16}
    \label{fig:res_chaosT2Sparse16}
\end{figure}

Finally, the reconstructions were compared qualitatively with the help of difference images and SSIM maps for Sparse 16 are shown in Fig.~\ref{fig:res_chaosT1inSparse16}~and~\ref{fig:res_chaosT2Sparse16} for T1 in-phase and T2-SPIR, respectively. Further qualitative comparisons for Sparse 8, as well as for T1 opposed-phase are shown in Appendix (Sec.~\ref{sec_appendix_mri}).

\begin{figure}[!htbp]
    \centering
    \includegraphics[width=0.48\textwidth]{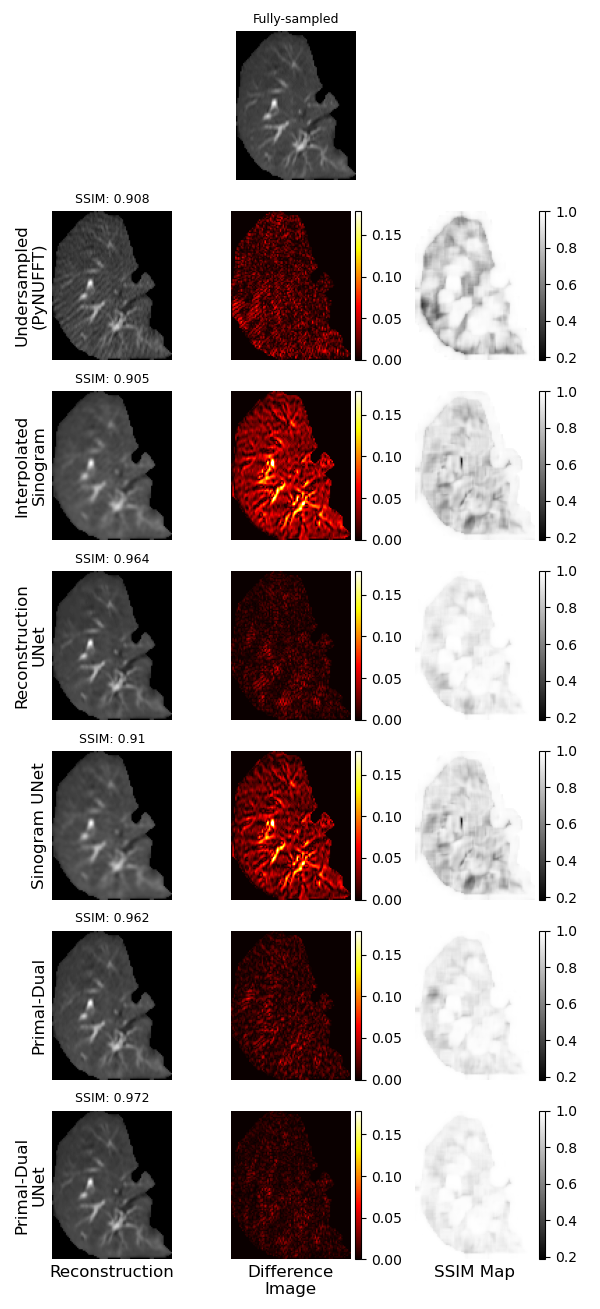}
    \caption{Qualitative comparisons of the reconstructions of CHAOS T2w Sparse 8: Liver ROI}
    \label{fig:res_chaosLiverROIT2}
\end{figure}

\paragraph{Evaluation for regions of interest (ROI)}
Further evaluations were performed to compare the performance of the proposed model against the other methods for different regions of interest. For this purpose, the images were segmented into three different regions: liver, kidneys (both left and right), and spleen, with the help of the available segmentation labels from the CHAOS dataset with the images. Then the images were cropped to have only the region of interest. Those segmented-cropped images obtained from the results of different methods were then compared against the segmented-cropped version of the ground-truth images. Tables~\ref{tab:res_CHAOS_kidneys},~\ref{tab:res_CHAOS_kidneys},~\ref{tab:res_CHAOS_spleen} show the quantitative results for liver, kidneys, and spleen. Statistical tests revealed that the proposed model archived statistically significant improvements over all the other methods. Fig.~\ref{fig:res_chaosLiverROIT2} shows a qualitative comparison of the reconstruction quality for liver, for Sparse 8.

\begin{table}[h!]
\caption{Resultant metrics for liver from CHAOS dataset (mean\textpm std)}
\label{tab:res_CHAOS_liver}
\resizebox{0.48\textwidth}{!}{%
\begin{tabular}{@{}lcccc@{}}
\toprule
 \multicolumn{1}{c}{\multirow{2}{*}{Method}}                           & \multicolumn{2}{c}{SSIM}                                        & \multicolumn{2}{c}{RMSE}                                         \\ \cmidrule(r){2-3}\cmidrule(l){4-5}
                      & \multicolumn{1}{c}{Sparse 8}             & Sparse 16            & \multicolumn{1}{c}{Sparse 8}             & Sparse 16            \\ \midrule
Undersampled (NUFFT)      & \multicolumn{1}{c}{0.831\textpm0.083}          & 0.656\textpm0.102          & \multicolumn{1}{c}{0.028\textpm0.014}          & 0.065\textpm0.026          \\
Bilinear Sinogram   & \multicolumn{1}{c}{0.883\textpm0.046}          & 0.788\textpm0.076          & \multicolumn{1}{c}{0.037\textpm0.016}          & 0.048\textpm0.021          \\
Sinogram UNet              & \multicolumn{1}{c}{0.893\textpm0.042}          & 0.826\textpm0.063          & \multicolumn{1}{c}{0.032\textpm0.014}          & 0.038\textpm0.016          \\
Reconstruction UNet        & \multicolumn{1}{c}{0.951\textpm0.029}          & 0.864\textpm0.061          & \multicolumn{1}{c}{0.013\textpm0.007}          & 0.026\textpm0.014          \\
Primal-Dual Network & \multicolumn{1}{c}{0.951\textpm0.030}          & 0.866\textpm0.059          & \multicolumn{1}{c}{0.012\textpm0.005}          & 0.025\textpm0.011          \\
\textbf{Primal-Dual UNet}  & \multicolumn{1}{c}{\textbf{0.963\textpm0.026}} & \textbf{0.888\textpm0.054} & \multicolumn{1}{c}{\textbf{0.011\textpm0.006}} & \textbf{0.021\textpm0.010} \\ \bottomrule
\end{tabular}%
}
\end{table}

\begin{table}[h!]
\caption{Resultant metrics for both left and right kidneys from CHAOS dataset (mean\textpm std)}
\label{tab:res_CHAOS_kidneys}
\resizebox{0.48\textwidth}{!}{%
\begin{tabular}{@{}lcccc@{}}
\toprule
 \multicolumn{1}{c}{\multirow{2}{*}{Method}}                           & \multicolumn{2}{c}{SSIM}                                        & \multicolumn{2}{c}{RMSE}                                         \\ \cmidrule(r){2-3}\cmidrule(l){4-5}
                      & \multicolumn{1}{c}{Sparse 8}             & Sparse 16            & \multicolumn{1}{c}{Sparse 8}             & Sparse 16            \\ \midrule
Undersampled (NUFFT)      & \multicolumn{1}{c}{0.877\textpm0.180}          & 0.695\textpm0.168          & \multicolumn{1}{c}{0.027\textpm0.014}          & 0.070\textpm0.028          \\
Bilinear Sinogram   & \multicolumn{1}{c}{0.866\textpm0.083}          & 0.759\textpm0.112          & \multicolumn{1}{c}{0.051\textpm0.023}          & 0.065\textpm0.031          \\
Sinogram UNet              & \multicolumn{1}{c}{0.881\textpm0.079}          & 0.809\textpm0.095          & \multicolumn{1}{c}{0.045\textpm0.020}          & 0.054\textpm0.025          \\
Reconstruction UNet        & \multicolumn{1}{c}{0.962\textpm0.054}          & 0.862\textpm0.086          & \multicolumn{1}{c}{0.016\textpm0.008}          & 0.035\textpm0.018          \\
Primal-Dual Network & \multicolumn{1}{c}{0.963\textpm0.057}          & 0.866\textpm0.077          & \multicolumn{1}{c}{0.016\textpm0.008}          & 0.039\textpm0.023          \\
\textbf{Primal-Dual UNet}  & \multicolumn{1}{c}{\textbf{0.971\textpm0.049}} & \textbf{0.892\textpm0.073} & \multicolumn{1}{c}{\textbf{0.013\textpm0.007}} & \textbf{0.030\textpm0.014} \\ \bottomrule
\end{tabular}%
}
\end{table}

\begin{table}[h!]
\caption{Resultant metrics for spleen from CHAOS dataset (mean\textpm std)}
\label{tab:res_CHAOS_spleen}
\resizebox{0.48\textwidth}{!}{%
\begin{tabular}{@{}lcccc@{}}
\toprule
 \multicolumn{1}{c}{\multirow{2}{*}{Method}}                           & \multicolumn{2}{c}{SSIM}                                        & \multicolumn{2}{c}{RMSE}                                         \\ \cmidrule(r){2-3}\cmidrule(l){4-5}
                      & \multicolumn{1}{c}{Sparse 8}             & Sparse 16            & \multicolumn{1}{c}{Sparse 8}             & Sparse 16            \\ \midrule
Undersampled (NUFFT)      & \multicolumn{1}{c}{0.848\textpm0.149}          & 0.699\textpm0.155          & \multicolumn{1}{c}{0.024\textpm0.010}          & 0.059\textpm0.021          \\
Bilinear Sinogram   & \multicolumn{1}{c}{0.878\textpm0.114}          & 0.817\textpm0.130          & \multicolumn{1}{c}{0.045\textpm0.018}          & 0.055\textpm0.022          \\
Sinogram UNet              & \multicolumn{1}{c}{0.885\textpm0.121}          & 0.840\textpm0.127          & \multicolumn{1}{c}{0.039\textpm0.017}          & 0.045\textpm0.019          \\
Reconstruction UNet        & \multicolumn{1}{c}{0.953\textpm0.104}          & 0.895\textpm0.117          & \multicolumn{1}{c}{0.011\textpm0.005}          & 0.023\textpm0.010          \\
Primal-Dual Network & \multicolumn{1}{c}{0.950\textpm0.106}          & 0.894\textpm0.116          & \multicolumn{1}{c}{0.012\textpm0.006}          & 0.025\textpm0.013          \\
\textbf{Primal-Dual UNet}  & \multicolumn{1}{c}{\textbf{0.957\textpm0.118}} & \textbf{0.907\textpm0.128} & \multicolumn{1}{c}{\textbf{0.009\textpm0.004}} & \textbf{0.019\textpm0.009} \\ \bottomrule
\end{tabular}%
}
\end{table}

\subsection{Comparison of Execution Speeds}
\begin{figure}[h!]
    \centering
    \includegraphics[width=0.45\textwidth]{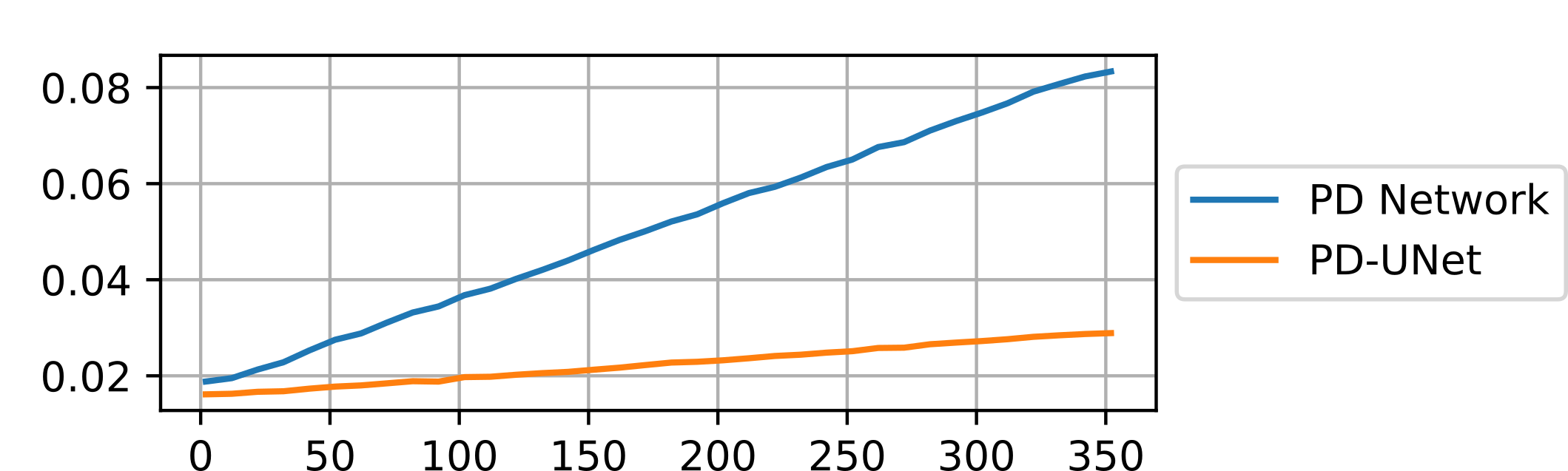}
    \caption{Comparison of inference times for Primal-Dual Network (PD Network) and Primal-Dual UNet (PD-UNet). Number of projections on horizontal axis. Average inference time for one batch (batch size: 4) in seconds on vertical axis.}
    \label{fig:speed}
\end{figure}

Sparse sampling can reduce the speed of acquisition, which is an essential factor when it comes to MR imaging. However, the time required for reconstruction can be an additional overhead - increasing the total time for imaging. For this reason, the execution speed of the proposed Primal-Dual UNet was compared against the main baseline model - Primal-Dual Network. Fig.~\ref{fig:speed} shows the required amount of time to reconstruct one slice for these two methods and how much they change with a change in the number of projections. It can be observed that the proposed method is faster than the baseline Primal-Dual Network. Moreover, it can be observed that with the increase in the number of projections, the required reconstruction time increases for both models, but also the difference between the models increases constantly. 

\subsection{Network Complexity}
\begin{table}
\caption{Network complexity: Analysis of parameters and forward pass memory (during inference) with a batch size of 1 and an image size of 256x256. ($BbCc$) encodes the number of blocks/iterations $b$ and the number of convolutional layers per block $c$.}
\label{tab:net-complex}
\resizebox{0.48\textwidth}{!}{%
\begin{tabular}{@{}lrl@{}}
\toprule
Network             & \#Parameters & Forward Pass in MB \\
                    &              & (Sparse 4/8/16)    \\
\midrule
PD-UNet             & 3,625,764            & 647/598/574                        \\
PD-Network ($B10C5$)  & 253,320              & 1,186/941/822                      \\
PD-Network ($B150C5$) & 3,799,800            & 17,799/14,129/12,335               \\
PD-Network ($B10C301$) & 3,675,080            & 3,827/3,038/2,652                 \\
\bottomrule
\end{tabular}
}
\end{table}
Tab.~\ref{tab:net-complex} shows a comparison of the proposed Primal-Dual UNet with the baseline Primal-Dual Network (in the table: $B10C5$) and variations of it with respect to the number of trainable parameters and the memory needed for a forward pass during inference. Although Primal-Dual UNet consists of almost fifteen times as many parameters as the baseline Primal-Dual Network, it only takes half the memory for inference. To make the networks more comparable in terms of parameters, two variations of the baseline are shown as well: For ($B150C5$), the number of blocks/iterations was increased from 10 to 150, resulting in a similar amount of parameters to Primal-Dual UNet. However, this increases the memory footprint for an inference to more than 12 GB, making it barely runnable, if not impossible to run, on current consumer GPUs. For the second variation ($B10C301$), the number of blocks was kept at 10, but the number of convolutional layers per block was increased to 301, resulting in a similar parameter count as Primal-Dual UNet. Again, the memory footprint for inference is increased compared to the baseline, however not exceeding 4 GB.

\section{Discussion}\label{sec:Discussion}

The results revealed that all four deep learning based models performed better than applying FBP on the bilinearly upsampled sinograms. Sinogram UNet, which aims to refine those interpolated sinograms before applying FBP, performed the worst among the deep learning models. For the lowest level of sparsity (Sparse 4), Reconstruction UNet resulted in the same average SSIM as the main baseline of this paper - Primal-Dual Network - but resulted in a better average RMSE. However, the superiority of the Primal-Dual Network can be seen for the higher levels of sparsity. For Sparse 4 and 16, the proposed Primal-Dual UNet performed better than the baseline Primal-Dual Network. However, interestingly, the results of Sparse 8 are conflicting for average SSIM and RMSE. According to the average RMSE, Primal-Dual UNet outperformed the Primal-Dual Network - in accordance with the other sparsity levels, but resulted in a lower average SSIM than the baseline. However, as the Primal-Dual UNet performed better in five out of six scenarios, it can be concluded as the overall better-performing model.

In the scope of CT, the needle insertion experiments showed that the networks not only learn to reconstruct images with tissues that they were trained on but also remove the artefacts from materials with significantly different (in this case: higher) attenuation coefficients without further training. As briefly described above, the volumes with the inserted needles were merely simulated and therefore lack other types of artefacts, e.g. caused by photon starvation (due to the high attenuation coefficients of the needles) or motion (due to the breathing of the subjects). Moreover, the patch size of 32x32 for the ROI evaluation around the needle was chosen empirically for this experiment to include the needle and some of the soft tissue which was affected most by the needle artefacts. Though this gives a good insight to how the different networks behave on unknown data, the quantitative values are susceptible to changes in the patch size and the shape of the depicted needle. Here, the shape was dot-like for the reconstructed axial planes, but it may as well be similar to a straight or bent line if inserted differently, which would invalidate the chosen patch size. In addition, conventional CT scanners nowadays usually use multirow detectors and helical acquisition trajectories, which was not taken care of in this study and might further increase the quality of the reconstructions.

Regarding CT-inspired techniques for performing radial MRI reconstruction, an interesting fact to be noted is that the first-ever 2D MRI was also produced using a back-projection algorithm~\citep{lauterbur1973image,geva2006magnetic}. The only difference between the sparse CT reconstruction and undersampled radial MRI reconstruction using the proposed model (as well as the CT-inspired baselines) is the 1D inverse Fourier transform as pre-processing. The experiments performed as a part of this research show the possibility of using sinogram upsampling techniques combined with FBP to reconstruct undersampled radial MRI.

It was observed that converting the undersampled radial k-space into the corresponding sinogram, then applying bilinear interpolation before finally performing FBP already results in scores better than the undersampled radial MRIs which were reconstructed with the conventional NUFFT. Sinogram UNet, which aims to improve the quality of the bilinearly upsampled sinograms, improves the results even further in terms of SSIM. Even though these two methods perform better than the traditional reconstruction of the undersampled radial MRIs using NUFFT, the quantitative and qualitative evaluations yield that they are much inferior to the other three models. It is to be  Reconstruction UNet - a well-established model for deep learning based undersampled MRI reconstruction, which is the only model out of the four models which is not CT-inspired and works directly with the reconstructed images, performed very similar to the main baseline model of this paper: Primal-Dual Network. Reconstruction UNet performed better for the brain MRI reconstruction task, whereas the Primal-Dual Network performed better while reconstructing abdominal MRIs. As noted earlier, the proposed method Primal-Dual UNet performed better than all the other methods for all the tasks. Another important point to note is that the trainings must be performed for each interesting organ and type of MRI (e.g. different MR contrasts), in line with the previously published research, and how it has also been done here: specific models for brain T1w MRI, abdominal T1w In-Phase, abdominal T1w Opposed-phase, and abdominal T2w MRIs. It might be possible to train a single model with different types of MRIs or organs - this is an interesting avenue to explore in the future.

The analysis of the network complexity has shown that Primal-Dual UNet's memory requirements are, in all cases, less than those of the baseline. Primal-Dual Network and its variations are not scalable and might exceed GPU memory limits for relatively small reconstruction problems already. Primal-Dual UNet, on the other hand, suffers much less from scalability problems due to the down-scaling/pooling in its UNet blocks, which makes it more likely to be used in practice and to be developed further in future research.

\section{Conclusion}\label{sec:Conclusion}
This paper presents a novel Primal-Dual UNet model for performing sinogram upsampling, which improves upon the Primal-Dual Network in terms of accuracy and speed of reconstruction. The method was employed for the task of sparse abdominal CT reconstruction for three different levels of sparsity: 4, 8, and 16. Moreover, this paper shows the potential of using sinogram upsampling techniques, such as deep learning based techniques, for radially-acquired undersampled MRI reconstruction. The proposed method was applied for the task of undersampled brain and abdominal MRI reconstruction with radial acquisitions for acceleration factors 8 and 16. The proposed method outperformed the baseline methods for most CT and all MRI experiments with statistical significance. Needle insertion simulation experiment showed that the proposed network not only yields better image reconstruction, but also generalises better than the other networks - reconstructs images with needle significantly better than the rest while being trained on a dataset without any needle. Moreover, region-of-interest based evaluations showed significant improvements over the baselines not only for the overall images, but also for the regions-of-interest liver, kidneys, and spleen.  

Future work will focus on applying the proposed method to cone-beam CT projections, thus yielding true 3D reconstructions, and incorporating known information about the volume for facilitating partial volume updates to further reduce memory and time requirements of the reconstruction task. Moreover, data augmentation techniques were only used minimally here, but could have a big impact on the generalisability and overall quality of the networks. Using data from different acceleration/sparsity factors during training could not only eliminate the need for one model per factor, but even help the network learn how subsampling the data influences the creation of artefacts and ultimately learn how to suppress these artefacts independent of the factor, which might make the model applicable to unseen subsampling patterns - will also be evaluated.

\section*{Acknowledgement}
This work was conducted within the context of the International Graduate School MEMoRIAL at Otto von Guericke
University (OVGU) Magdeburg, Germany, kindly supported by the European Structural and Investment Funds (ESF) under the
programme ``Sachsen-Anhalt WISSENSCHAFT Internationalisierung'' (project no. ZS/2016/08/80646).

\bibliography{mybibfile}

\newpage
\section*{Appendix}

\subsection{Reconstruction of CT images with parallel-beam geometry}
\label{sec_appendix_parallel}

\begin{figure}[h!]
    \centering
    \includegraphics[width=0.48\textwidth]{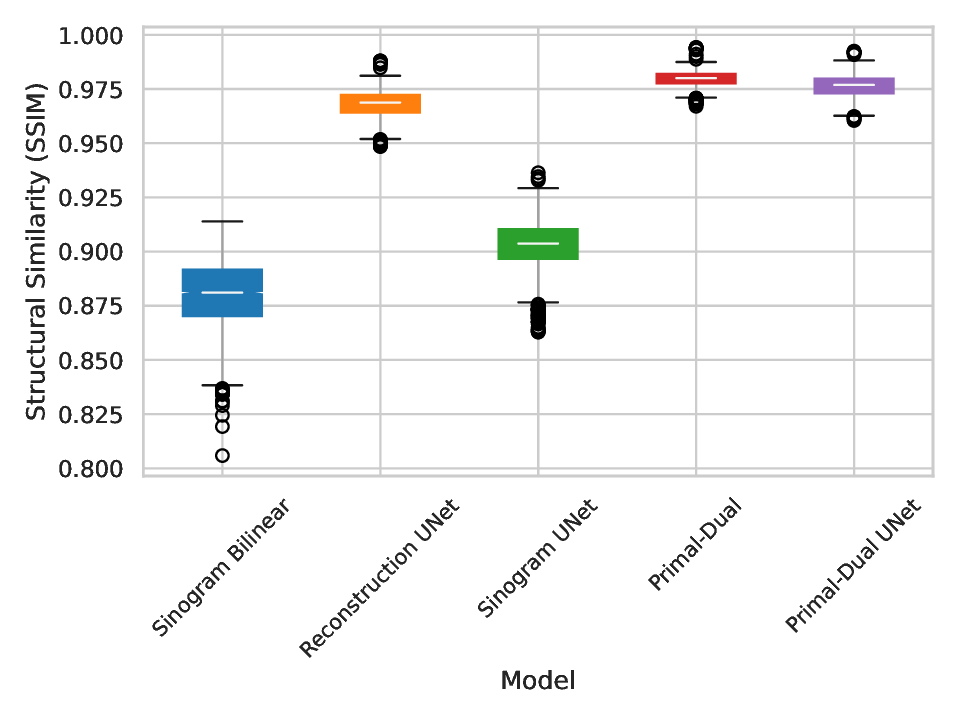}
    \caption{Box-plots of the resultant SSIM values for CT (parallel-beam geometry) Sparse 4}
    \label{fig:box_ct_par_Sparse4}
\end{figure}

\begin{figure}[h!]
    \centering
    \includegraphics[width=0.48\textwidth]{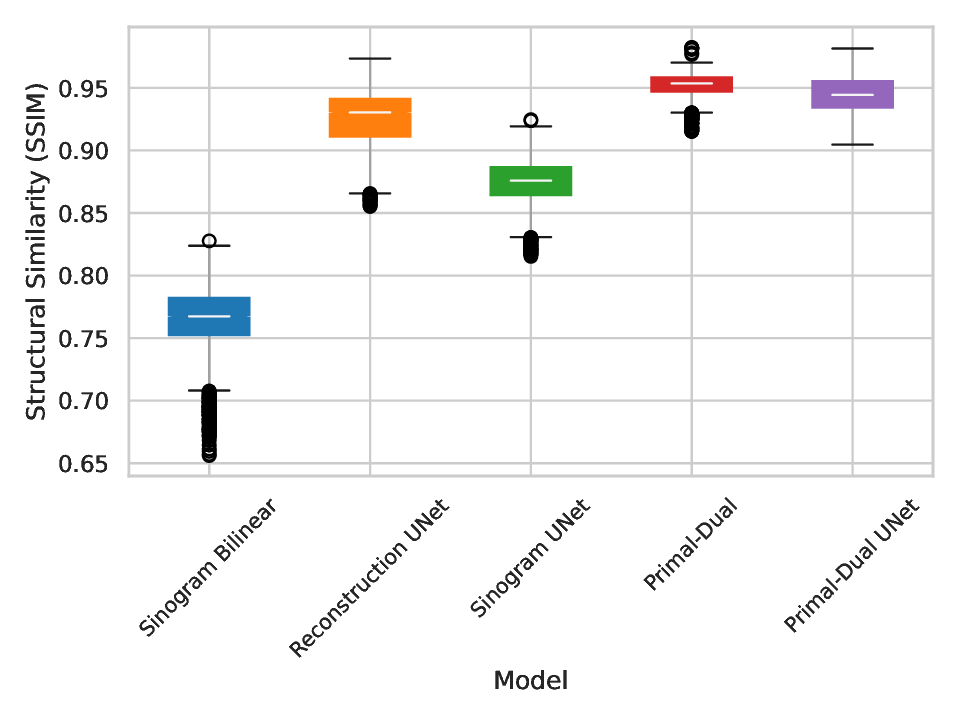}
    \caption{Box-plots of the resultant SSIM values for CT (parallel-beam geometry) Sparse 8}
    \label{fig:box_ct_par_Sparse8}
\end{figure}

\begin{figure}[h!]
    \centering
    \includegraphics[width=0.48\textwidth]{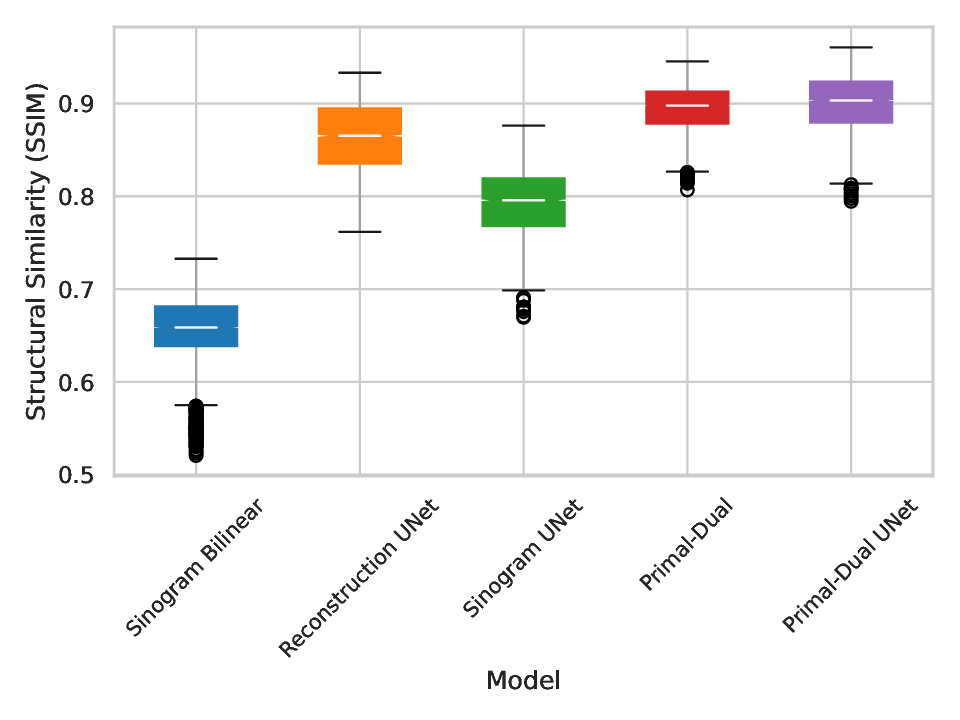}
    \caption{Box-plots of the resultant SSIM values for CT (parallel-beam geometry) Sparse 16}
    \label{fig:box_ct_par_Sparse16}
\end{figure}

\begin{figure}
    \centering
    \includegraphics[width=0.48\textwidth]{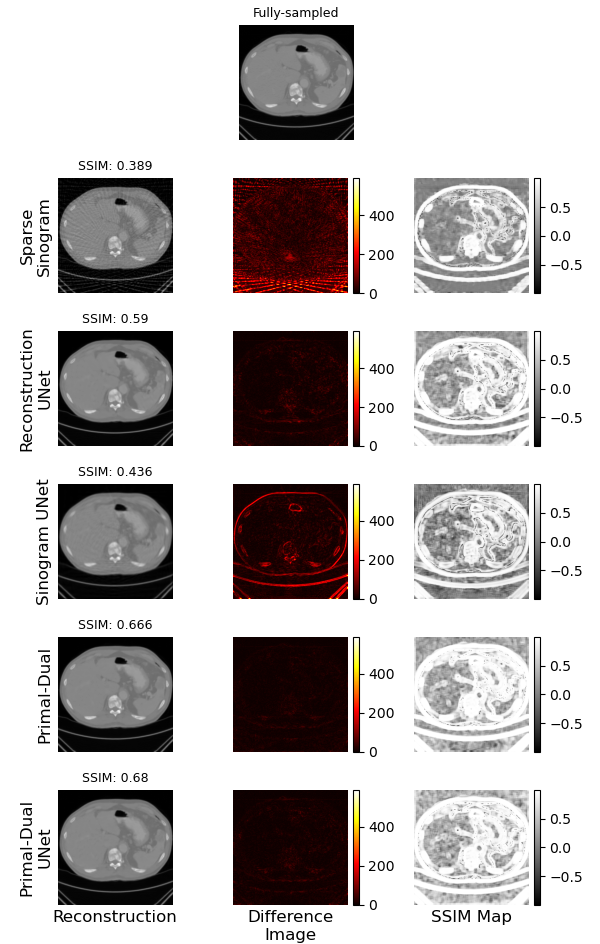}
    \caption{Qualitative comparisons of the reconstructions of CT (parallel-beam geometry) Sparse 4}
    \label{fig:res_ct_par_Sparse4}
\end{figure}

\begin{figure}
    \centering
    \includegraphics[width=0.48\textwidth]{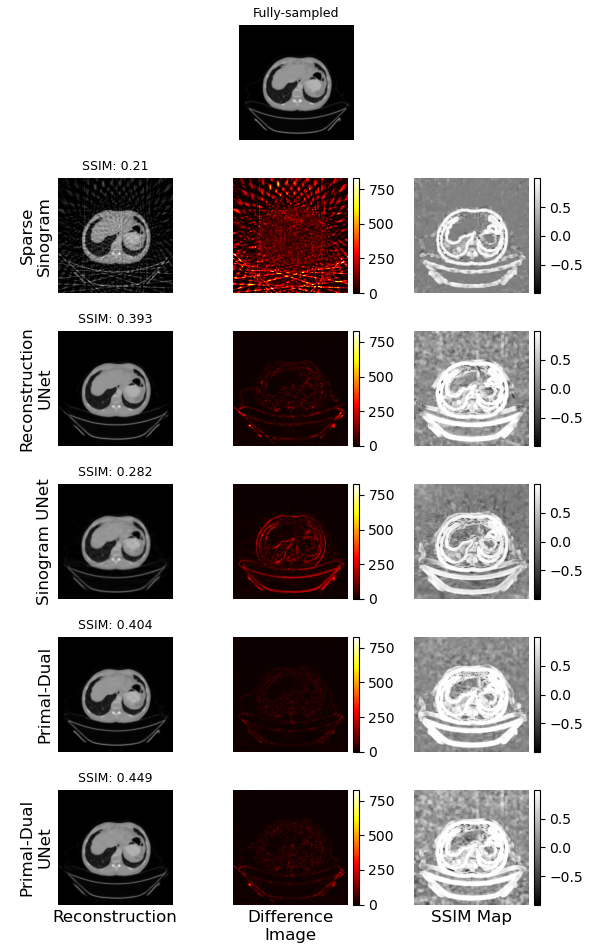}
    \caption{Qualitative comparisons of the reconstructions of CT (parallel-beam geometry) Sparse 8}
    \label{fig:res_ct_par_Sparse8}
\end{figure}

\begin{figure}
    \centering
    \includegraphics[width=0.48\textwidth]{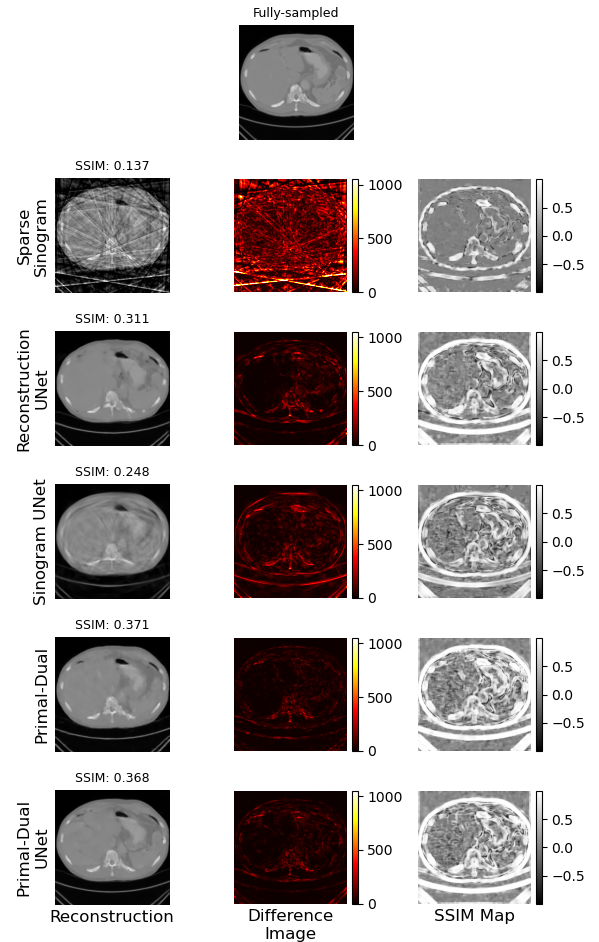}
    \caption{Qualitative comparisons of the reconstructions of CT (parallel-beam geometry) Sparse 16}
    \label{fig:res_ct_par_Sparse16}
\end{figure}

This section presents the results for the sparse CT reconstruction for the parallel-beam geometry. Each projection consists of 363 detector pixels with a pixel spacing of \SI{1}{px} from parallel-beams to cover the full axial slice. The sinograms contain 180 equiangular projections with an angular distance of \ang{1} between consecutive projections. Tab.~\ref{tab:res_CT_par} shows the quantitative comparison of the methods in terms of SSIM (calculated on the normalised intensity values) and RMSE (in the Hounsfield scale), and Fig.~\ref{fig:box_ct_par_Sparse4}, \ref{fig:box_ct_par_Sparse8}, and \ref{fig:box_ct_par_Sparse16} portray the range of the resultant SSIM values for the three levels of sparsity: 4, 8, and 16, respectively. In terms of SSIM, the baseline Primal-Dual Network performed better than the proposed method for Sparse 4 and 8. However, the proposed method outperformed the baseline Primal-Dual Network for the highest level of sparsity experimented here: 16. Regarding RMSE, the baseline Primal-Dual Network performed better than the proposed method. However, the proposed method performed better than the other baseline methods - in terms of both, SSIM and RMSE. Finally, qualitative comparisons  of the reconstructions with the help of difference images and SSIM maps are shown in Fig.~\ref{fig:res_ct_par_Sparse4}, \ref{fig:res_ct_par_Sparse8}, and \ref{fig:res_ct_par_Sparse16}, for Sparse 4, 8, and 16, respectively. 

\begin{table*}[]
\centering
\caption{Resultant metrics for the CT parallel-beam geometry}
\label{tab:res_CT_par}
\begin{tabular}{@{}lcccccc@{}}
\toprule
\multicolumn{1}{c}{\multirow{2}{*}{Method}}                           &
  \multicolumn{3}{c}{SSIM} &
  \multicolumn{3}{c}{RMSE (HU)} \\ \cmidrule(r){2-4}\cmidrule(l){5-7} 
 &
  \multicolumn{1}{c}{Sparse 4} &
  \multicolumn{1}{c}{Sparse 8} &
  Sparse 16 &
  \multicolumn{1}{c}{Sparse 4} &
  \multicolumn{1}{c}{Sparse 8} &
  Sparse 16 \\ \midrule
Bilinear Sinogram &
  \multicolumn{1}{c}{0.880\textpm0.015} &
  \multicolumn{1}{c}{0.764\textpm0.028} &
  0.654\textpm0.038 &
  \multicolumn{1}{c}{42.504\textpm5.337} &
  \multicolumn{1}{c}{72.237\textpm9.285} &
  109.353\textpm13.098 \\
Sinogram UNet &
  \multicolumn{1}{c}{0.903\textpm0.011} &
  \multicolumn{1}{c}{0.875\textpm0.019} &
  0.793\textpm0.037 &
  \multicolumn{1}{c}{50.024\textpm3.921} &
  \multicolumn{1}{c}{53.618\textpm5.887} &
  70.777\textpm11.304 \\
Reconstruction UNet &
  \multicolumn{1}{c}{0.968\textpm0.006} &
  \multicolumn{1}{c}{0.925\textpm0.022} &
  0.863\textpm0.037 &
  \multicolumn{1}{c}{17.806\textpm2.964} &
  \multicolumn{1}{c}{35.180\textpm10.355} &
  62.507\textpm19.468 \\
Primal-Dual Network &
  \multicolumn{1}{c}{\textbf{0.980\textpm0.003}} &
  \multicolumn{1}{c}{\textbf{0.952\textpm0.010}} &
  0.895\textpm0.026 &
  \multicolumn{1}{c}{\textbf{11.913\textpm1.660}} &
  \multicolumn{1}{c}{\textbf{21.456\textpm4.212}} &
  \textbf{43.926\textpm13.337} \\
\textbf{Primal-Dual UNet} &
  \multicolumn{1}{c}{0.976\textpm0.005} &
  \multicolumn{1}{c}{0.944\textpm0.014} &
  \textbf{0.899\textpm0.032} &
  \multicolumn{1}{c}{14.579\textpm2.276} &
  \multicolumn{1}{c}{26.545\textpm6.151} &
  46.044\textpm13.362 \\ \bottomrule
\end{tabular}%
\end{table*}

\FloatBarrier

\subsection{Qualitative results for CT reconstructions with fan-beam geometry for Sparse 4}
\label{sec_appendix_fan_4}
\begin{figure}[h!]
    \centering
    \includegraphics[width=0.48\textwidth]{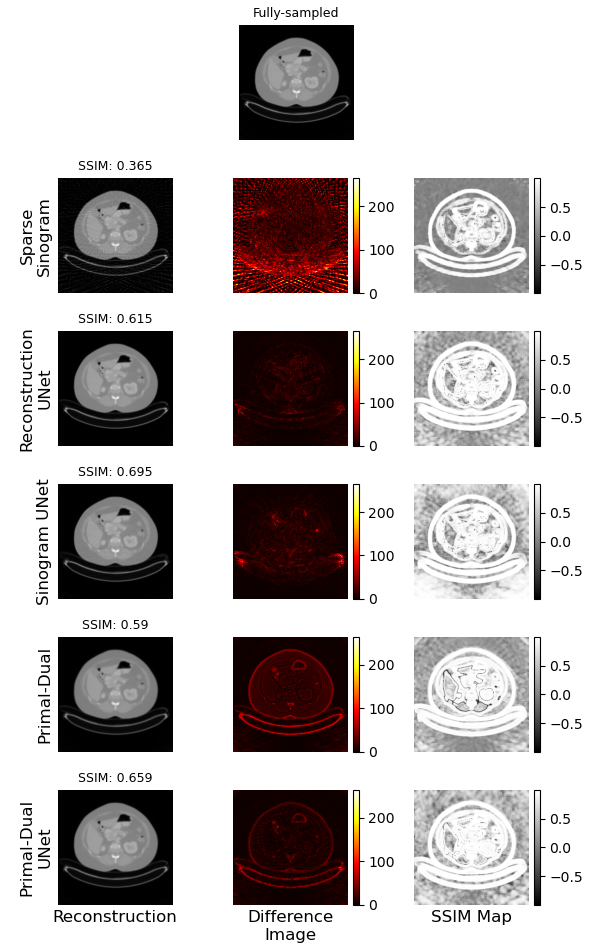}
    \caption{Qualitative comparisons of the reconstructions of CT (fan-beam geometry) Sparse 4}
    \label{fig:res_ct_fan_Sparse4}
\end{figure}

The qualitative comparison for reconstructions of CT images with sparsity level for four for the fan-beam geometry is shown here, in Fig.~\ref{fig:res_ct_fan_Sparse4}.

\newpage

\subsection{Qualitative results for MRI reconstructions}
\label{sec_appendix_mri}
\begin{figure}[h!]
    \centering
    \includegraphics[width=0.48\textwidth]{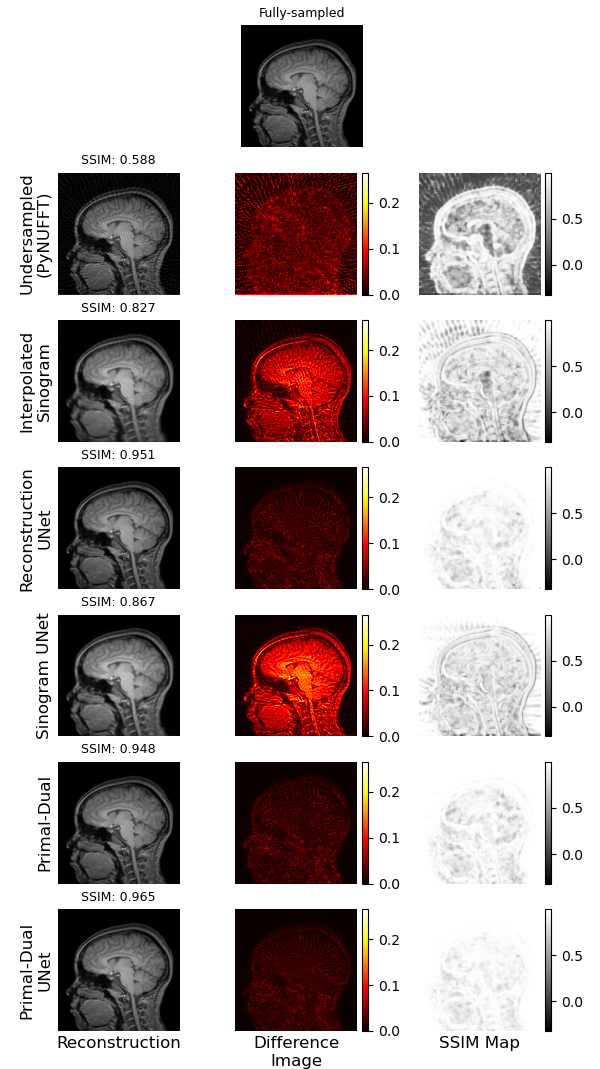}
    \caption{Qualitative comparisons of the reconstructions of MRI Sparse 8 for the IXI dataset}
    \label{fig:res_ixiSparse8}
\end{figure}

\begin{figure}[h!]
    \centering
    \includegraphics[width=0.48\textwidth]{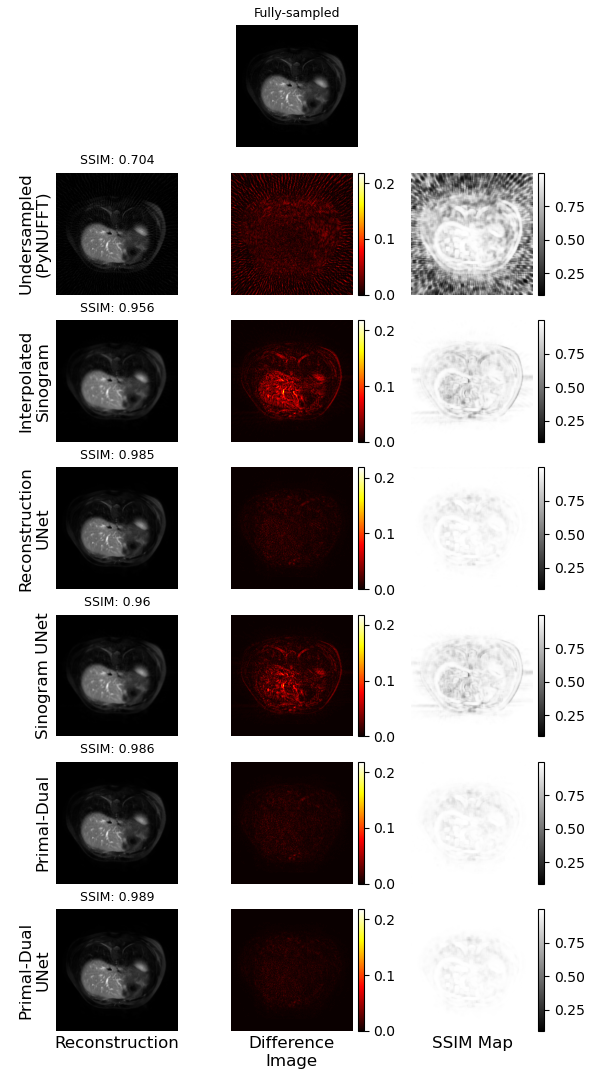}
    \caption{Qualitative comparisons of the reconstructions of CHAOS T2w Sparse 8}
    \label{fig:res_chaosT2Sparse8}
\end{figure}

\begin{figure}[h!]
    \centering
    \includegraphics[width=0.48\textwidth]{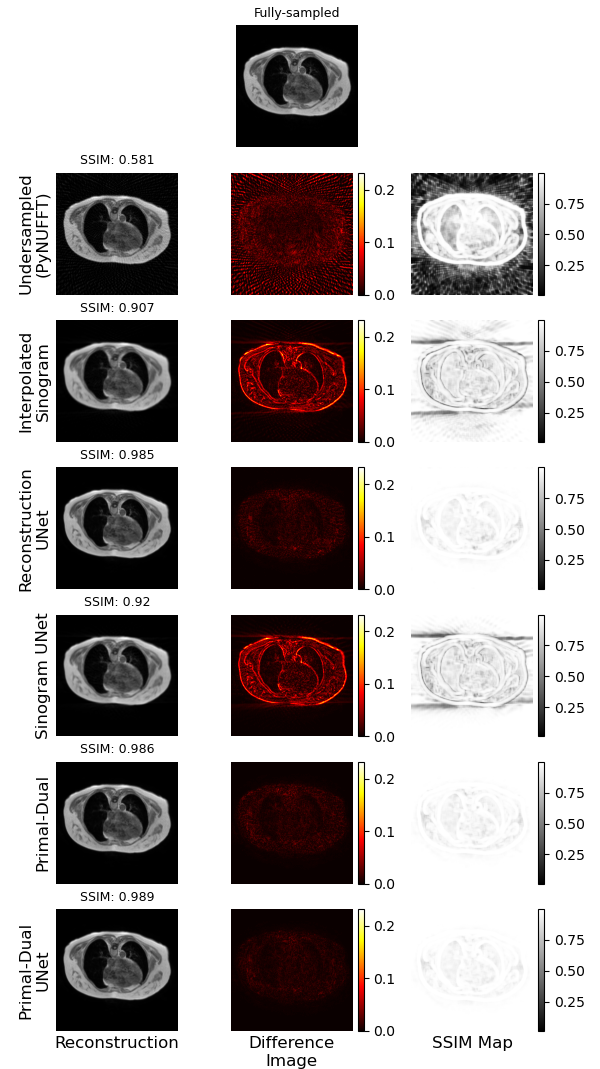}
    \caption{Qualitative comparisons of the reconstructions of CHAOS T1w Opposed-phase Sparse 8}
    \label{fig:res_chaosT1outSparse8}
\end{figure}

\begin{figure}[h!]
    \centering
    \includegraphics[width=0.48\textwidth]{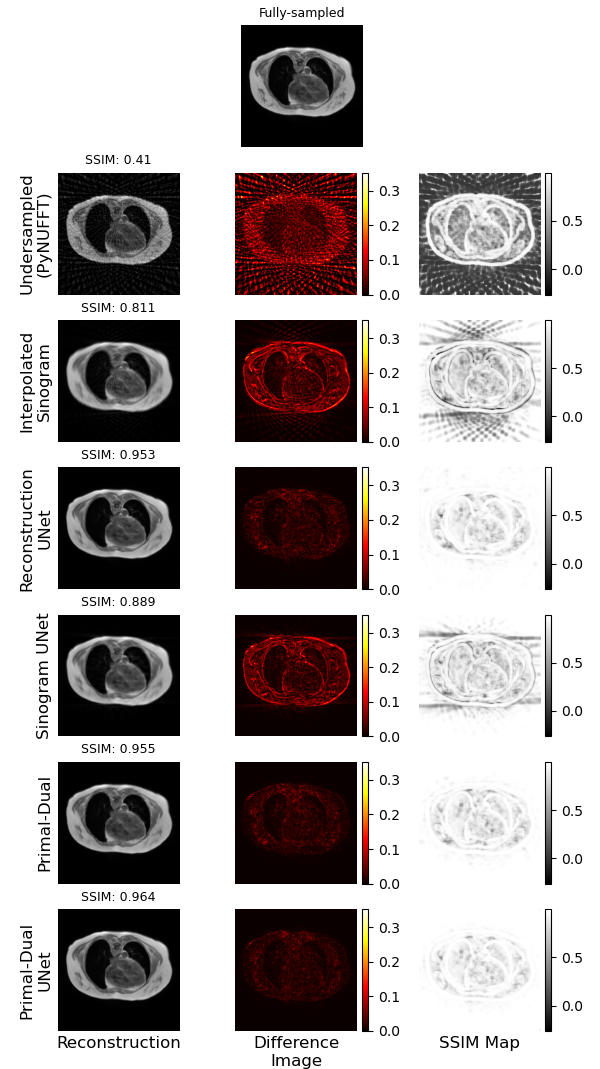}
    \caption{Qualitative comparisons of the reconstructions of CHAOS T1w Opposed-phase Sparse 16}
    \label{fig:res_chaosT1outSparse16}
\end{figure}

In this section, the qualitative comparisons for sparse 8 reconstructions are shown in Figs.~\ref{fig:res_ixiSparse8}, \ref{fig:res_chaosT2Sparse8}, and \ref{fig:res_chaosT1outSparse8} - for IXI, CHAOS T2-SPIR, and CHAOS T1 In-Phase, respectively. Finally, Figs.~\ref{fig:res_chaosT1outSparse8}~and~
\ref{fig:res_chaosT1outSparse16} portray the qualitative comparisons for CHAOS T1 Opposed-Phase, for Sparse 8 and 16, respectively. 

\begin{figure}[h!]
    \centering
    \includegraphics[width=0.48\textwidth]{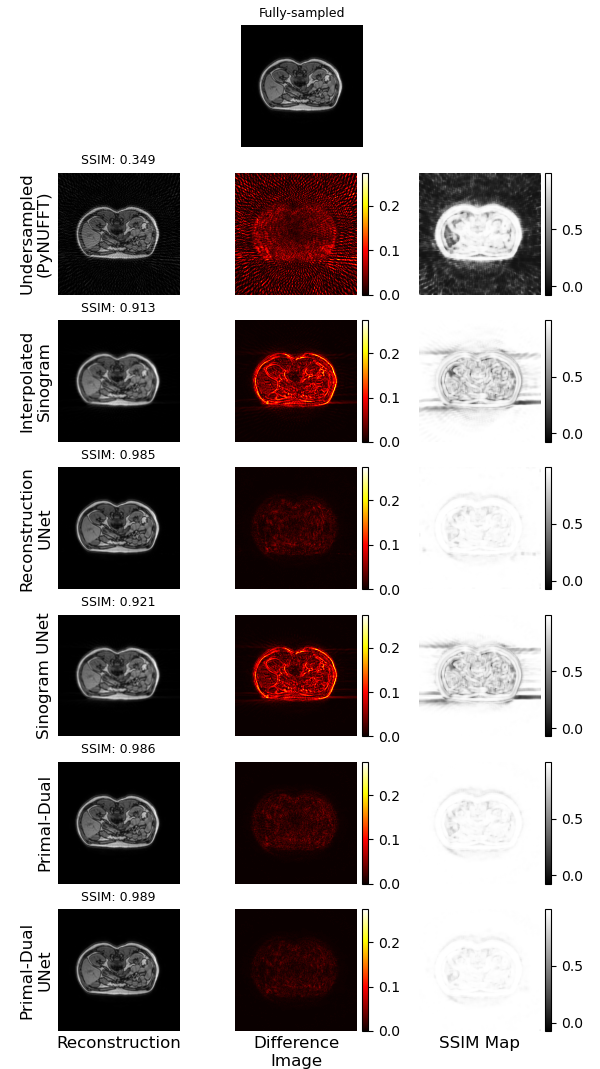}
    \caption{Qualitative comparisons of the reconstructions of CHAOS T1w In-phase Sparse 8}
    \label{fig:res_chaosT1inSparse8}
\end{figure}

\end{document}